\documentclass[a4paper,11pt]{article}
\pdfoutput=1 

\usepackage{jcappub}

\usepackage[T1]{fontenc} 

\usepackage{graphicx}
\usepackage{amsmath,amssymb}
\usepackage{braket}
\usepackage{subfigure}
\usepackage{float}
\usepackage[dvipsnames]{xcolor}
\usepackage{mathrsfs}
\usepackage{comment}
\usepackage{multirow}
\usepackage{soul}
\usepackage{mathtools}
\usepackage{sidecap} 
\usepackage{enumitem}
\graphicspath{{Figures/}}
\usepackage[normalem]{ulem}
\setlist{itemsep=2pt,topsep=4pt}
\graphicspath{{figures/}}
\numberwithin{equation}{section}
\usepackage[dvipsnames]{xcolor}
\usepackage{bm}
\usepackage{booktabs}
\usepackage{array}
\usepackage{tcolorbox}

\newcommand{\be}{\begin{equation}}
\newcommand{\ee}{\end{equation}}

\newcommand{\ii}{\mathrm{i}}
\newcommand{\dd}{\mathrm{d}}
\newcommand{\e}{\mathrm{e}}
\newcommand{\bx}{\bm{x}}
\newcommand{\by}{\bm{y}}
\newcommand{\bn}{\hat{\bm n}}
\newcommand{\bp}{\bm{p}}
\newcommand{\bu}{\bm{u}}

\newcommand{\hF}{{}_1F_1}
\newcommand{\Tf}{\mathcal{T}}
\newcommand{\rh}{\hat\rho}
\newcommand{\tauh}{\hat\tau}
\newcommand{\Lc}{\mathcal{L}}
\newcommand{\Eosc}{E_{\rm osc}}
\newcommand{\Acl}{A_{\rm cl}}
\newcommand{\tdk}{\chi}
\newcommand{\gb}[1]{\textcolor{Emerald}{GB: #1}}

\newtcolorbox{keybox}[1][]{breakable,colback=blue!3!white,colframe=blue!45!black,boxrule=0.5pt,arc=2pt,
  left=6pt,right=6pt,top=4pt,bottom=4pt,fonttitle=\bfseries,#1}
\newtcolorbox{physbox}[1][]{breakable,colback=orange!4!white,colframe=orange!60!black,boxrule=0.5pt,arc=2pt,
  left=6pt,right=6pt,top=4pt,bottom=4pt,fonttitle=\bfseries,#1}

\title{Gravitational-wave lensing beyond the diffraction integral:\\
proper-time methods and the bound lens}
\author[a,b]{Ginevra Braga,}
\affiliation[a]{Gran Sasso Science Institute, Viale F. Crispi 7, I-67100 L'Aquila, Italy}
\affiliation[b]{INFN-Laboratori Nazionali del Gran Sasso, Via G. Acitelli 22, 67100 Assergi (AQ), Italy}
\author[c,d]{Raul Jimenez,}
\affiliation[c]{ICC, University of Barcelona, Mart\'i i Franqu\`es 1, E08028 Barcelona, Spain}
\affiliation[d]{ICREA, Pg. Llu\'is Companys 23, 08010 Barcelona, Spain}

\emailAdd{ginevra.braga@gssi.it}
\emailAdd{raul.jimenez@icc.ub.edu}

\abstract{
Gravitational waves (GWs) can have wavelengths comparable to the Schwarzschild radius of their lenses, and detectors
record the waveform, phase included, so their lensing must be described by wave optics. Its standard tool, the
diffraction integral, follows from writing the scalar wave on a static, weak-field background as a particle path
integral, under the eikonal (one-way) and paraxial approximations, and then imposing the thin-lens limit. We give a
self-contained account of how the proper-time (worldline) representation of the propagator goes beyond these steps. An
auxiliary proper time turns the wave equation into an exact first-order one, with the frequency playing the role of
$1/\hbar$, so that geometric optics and, for paraxial paths through a thin lens, the diffraction integral follow from
stationarity conditions of an exact integral. For a point mass, the Duru--Kleinert gauge and the Kustaanheimo--Stiefel
map give the propagator exactly; for an observer at infinity the Duru--Kleinert time and the fibre angle become the
radius and the azimuth of a lens plane, and geometric optics requires $w\gg1$ and $wy\gg1$. We also evaluate the eikonal
condition exactly, $\varepsilon_E\simeq\bar\lambda/d+GM/c^2d+\mathcal O(\sin^2\theta/2)$, and show that the one-way step
is a non-relativistic limit. As an application we treat the {\it bound lens}, a GW source orbiting its lens, which
emits a backscattered echo and moves during the lensing delays, so that all these approximations fail together, and we
derive its exact time-domain signal, in which each worldline carries its own retarded time. In the diffraction regime a
single straight worldline carrying the Shapiro phase misses all lensing beyond the Shapiro delay, geometric optics
misses 46--100\% of it near conjunction at $w=0.3$, the paraxial geometry fails on the far side of the orbit, and the
static amplification factor evaluated along the orbit errs by $\approx(1$--$2)\,wM/R$ near conjunction and by the echo
amplitude elsewhere.
}

\begin{document}
\maketitle

\section{Introduction}\label{sec:intro}

With more than two hundred gravitational-wave (GW) events observed by the LIGO--Virgo--KAGRA
collaboration~\cite{LIGOScientific:2018mvr,LIGOScientific:2020ibl,KAGRA:2021vkt,LIGOScientific:2025slb}, and with the
space interferometer LISA~\cite{amaro2017laser,LISA:2024hlh} and the third-generation ground-based
detectors~\cite{Punturo:2010zz,Reitze:2019iox,Branchesi:2023mws} in preparation, GWs have become a probe not only of
their sources but also of the matter they cross on their way to us. Gravitational lensing is one of the most important phenomena in which this is manifest: as
for light, the masses along the line of sight deflect, delay and focus the waves~\cite{SEF1992,Grespan:2023cpa}.
Unlike light, however, GWs can have wavelengths comparable to the Schwarzschild radius of the lens.The ratio of the two is measured by the dimensionless frequency $w=4GM_z\omega/c^3$, where $M_z$ is the redshifted
mass of the lens. Rays exist only for $w\gg1$, which defines the geometric-optics (GO)
regime~\cite{Isaacson:1967sln,Isaacson1968GravitationalRI}. When $w$ is of order unity, instead, the wave diffracts
around the lens and lensing becomes frequency dependent: this is the wave-optics
regime~\cite{Ohanian1974,Bontz:1981rvr,DeguchiWatson1986,Nakamura1998,NakamuraDeguchi1999,TakahashiNakamura2003}. Which lenses fall in this regime depends on the observed frequency. In the
band of ground-based detectors, $f\sim10$--$10^3$\,Hz, $w=1$ corresponds to lens masses of $8$--$800\,M_\odot$, i.e.\
stars and stellar or intermediate-mass black holes, while in the LISA band, $f\sim10^{-4}$--$10^{-1}$\,Hz, it
corresponds to $8\times10^4$--$8\times10^7\,M_\odot$, the masses of the black holes in galactic nuclei~\cite{Ohanian1974,Bontz:1981rvr,DeguchiWatson1986,Nakamura1998,NakamuraDeguchi1999,TakahashiNakamura2003}.
Moreover, interferometers record the strain itself, phase included, so that the complex amplification factor $F(f)$
multiplying the unlensed spectrum is observable, together with the relative phases of the images~\cite{Dai2017,Ezquiaga2021},
whereas in optical astronomy only $|F|^2$ is measured and interference is washed out by the size of the
source~\cite{Leung2023}. Wave-optics signatures are expected to be detectable with present and future
interferometers~\cite{TakahashiNakamura2003,Gao:2021sxw,Dai:2018enj,Cheung:2020okf}, can lift degeneracies between source
and lens parameters~\cite{Cremonese:2021puh} and can probe low-mass halos and compact dark objects~\cite{Oguri:2020ldf},
and the event GW231123~\cite{LIGOScientific:2025rsn} has already been analysed as a possible candidate for wave-optics
lensing~\cite{Goyal:2025eqo,Chan:2025kyu,Hu:2025lhv,Chakraborty:2025pxt}.

The standard description of wave-optics lensing was firstly introduce in the pioneering works of Nakamura and Deguchi~\cite{Nakamura1998,NakamuraDeguchi1999}
(see also~\cite{TakahashiNakamura2003,Takahashi2004,Leung2023}). Starting from the scalar wave equation in the weak
field of a static lens, one reduces the equation for the amplification factor to a Schr\"odinger equation in which the
distance from the source plays the role of time and the inverse frequency that of $\hbar$. Its solution is a path
integral over trajectories, as in quantum mechanics rather than over field configurations, an instance of the Hamilton
analogy between optics and mechanics~\cite{goldstein2002classical,Masoliver_2009}, and for a thin lens it collapses onto
the Kirchhoff--Fresnel \emph{diffraction integral} over the points of a lens plane, each weighted by the arrival time of
the path that crosses it~\cite{SEF1992,NakamuraDeguchi1999}. The same integral follows from a particle Hamiltonian built
on the dispersion relation~\cite{FPT2019}; it has a closed form for a point
mass~\cite{Peters1974,DeguchiWatson1986,Nakamura1998}, it is computed numerically for realistic
lenses~\cite{UlmerGoodman1995,Tambalo2023,GLoW2025}, and Picard--Lefschetz methods make it tractable at all
frequencies~\cite{FPT2019,Jow2023}. At high frequency only the stationary points of the arrival time survive, and one
recovers the rays, Fermat's principle and the laws of geometric optics.

This derivation rests on a chain of approximations: the \emph{eikonal} step, which reduces the second-order wave
equation to a first-order, one-way evolution; the \emph{paraxial} geometry, which replaces the sphere of directions by a
plane; a thin lens with source and observer in its far field; a weak field; a static configuration, without which no
amplification factor exists; and the scalar treatment of a tensor wave. For a lens at cosmological distance from both
source and observer they hold extremely well, but they are not universal, and their physical content is not always
transparent. The eikonal step in particular is easily confused with geometric optics, also because the word has several meanings in
the literature, although, as we show, it is a different and much weaker condition. Several recent works have looked beyond parts of this chain, with corrections beyond the eikonal
approximation~\cite{Bruyere:2026gnt}, a scattering-theory view of lensing~\cite{Takahashi:2005sxa,CarrilloGonzalez:2025gqm},
three-dimensional wave optics beyond the thin lens~\cite{Suyama2005,Carrillo-Gonzalez:2026oov}, wave optics in modified gravity, where the amplification factor no longer tends to unity at low
frequency~\cite{Garoffolo:2026hyx}, polarization
effects~\cite{Cusin:2019rmt,Dalang:2021qhu,Oancea2024}, rotating lenses~\cite{Bonga:2024orc} and the exact wave equation on a
Schwarzschild background~\cite{Pijnenburg2024}. When the wave crosses many lenses, the single-lens description gives way to a
statistical one: the amplitude and phase fluctuations produced by a random distribution of weak lenses have been computed in
wave optics~\cite{Macquart2004,Takahashi2006,MizunoSuyama2023}, and the problem has recently been formulated as a
Schwinger--Keldysh path integral in which the lens distribution acts as a quenched disorder, giving a single framework for
interference, diffraction and the loss of coherence~\cite{Amoruso:2026txw}.
A natural framework in which all these steps can be followed one at a time is the proper-time technique, or worldline
formalism. Introduced by Fock, Nambu, Feynman and Schwinger~\cite{Fock1937,Nambu1950,Feynman1950,Schwinger1951}, it
writes the propagator of a relativistic field as a sum over the trajectories of an associated particle, parametrized by
an auxiliary proper time that is integrated over at the end. It is the basis of the worldline formalism of quantum
field theory~\cite{Strassler1992,Schubert2001,BastianelliVN2006,Bastianelli:2013tsa,EdwardsSchubert2019}, in which the
proper time is the gauge-invariant length of the einbein of the
worldline~\cite{1976PhLB...64..435B,Brink:1976zz,Teitelboim1982,CMNP1986,Polyakov1987,HenneauxTeitelboim1992}; it has been applied to
wave propagation in optics and acoustics~\cite{FishmanMcCoy1984,Palmer1979,Schulman1984}, and in its
worldline-quantum-field-theory form it has recently become a tool for the gravitational two-body
problem~\cite{Mogull2021}. In Ref.~\cite{Braga2024} it was brought into GW lensing. The proper time turns the Helmholtz
equation \emph{exactly} into a first-order Schr\"odinger equation in which the frequency multiplies the whole worldline
action and plays the role of $1/\hbar$, with no assumption on the frequency, the direction of propagation, the size of
the lens or the scattering angle. The price is one more integral, over the proper time, which carries much of the
physics. The reward is that every approximation is taken \emph{a posteriori}, as a stationary point of an exact
integral, so that its validity can be judged: stationarity in the proper time alone, with paraxial and thin-lens paths,
gives the diffraction integral, and stationarity also in the path gives the rays. Both are semiclassical statements in
the strict sense, $\hbar=1/\omega\to0$.

Our aims are two. The first is to solve the point-mass lens exactly, beyond the diffraction integral. Its
proper-time path integral is a Coulomb problem, which we solve with the method of Duru and
Kleinert~\cite{DuruKleinert1979,DuruKleinert1982,Kleinert2009}, a change of clock along the worldline followed by the
Kustaanheimo--Stiefel map~\cite{KS1965}. The result is an exact representation of the Green function, in which the
lens-plane variables and the conditions for geometric optics acquire a direct worldline meaning. The second is to apply
this machinery to a configuration in which the approximations behind the diffraction integral fail together: the
\emph{bound lens}, a compact binary orbiting a massive black hole~\cite{Bellovary:2015ifg,Grishin2024,Meiron2017,Sberna2022,Kocsis2013},
whose repeated lensing has been proposed as a probe of such
systems~\cite{DOrazioLoeb2020,Yu2021,Pijnenburg2024,Cardoso:2026ugm,Ubach:2025dob}. There the source is close to the lens,
sweeps all angles around it, emits a backscattered echo and moves during the lensing delays. We compute its signal
exactly and use it as a benchmark for the approximations used in the literature.

\bigskip
\noindent
The paper is organized as follows. In Section~\ref{sec:standard} we review the standard route to the diffraction
integral, its geometric-optics limit, the point-mass solution, the Fresnel scale and the time-domain description in
terms of the delay density. In Section~\ref{sec:propertime} we introduce the proper-time technique,
Eq.~\eqref{eq:schwinger}, the role of the einbein and of the choice of clock, and the stationarity conditions that lead
to the diffraction integral and to geometric optics. In Section~\ref{sec:DK} we solve the point-mass problem exactly in
the Duru--Kleinert gauge, analyse its stationary contributions and show that, for an observer at infinity, the lens
plane is spanned by the DK time and the fibre angle. In Section~\ref{sec:bound} we discuss which assumptions of the
diffraction integral fail and why, including the exact eikonal condition, the non-relativistic meaning of the one-way
step and the backward waves, and we introduce the bound lens, computing the exact signal of a static and of a moving
source. In Section~\ref{sec:bound_models} we derive the approximate models of the literature from the exact formula and
define the lensing modulation of a periodic source, and in Section~\ref{sec:results} we measure their errors near
conjunction and on the far side of the orbit, estimate the corrections beyond the scalar linearized model, and discuss
chirping sources, distinguishability and the astrophysical window. We conclude in Section~\ref{sec:concl}. Technical
details are collected in the appendices. We use units $G=c=1$ unless stated otherwise.
\section{Standard wave optics: the diffraction integral}\label{sec:standard}

This section reviews the standard route to the diffraction integral \cite{NakamuraDeguchi1999,
TakahashiNakamura2003,Leung2023,Oguri:2020ldf,Takahashi:2005sxa}. We consider a static lens in the weak-field (Newtonian) metric
\begin{equation}\label{eq:metric}
\dd s^2=-\big(1+2U(\bx)\big)\dd t^2+\big(1-2U(\bx)\big)\dd\bx^2,\qquad |U|\ll1,
\end{equation}
where $U$ is the Newtonian potential. The standard treatment assumes that the GW polarization is parallel-transported along the rays and decouples from the propagation
\cite{Cusin:2019rmt} also in the wave optics regime. Hence one considers a 
massless scalar field $\varphi$ obeying $\Box\varphi=0$ on the background~\eqref{eq:metric}.
Because the lens is static, frequency is conserved and we may work with one Fourier component,
$\varphi=\tilde\varphi(\bx)\,\e^{-\ii\omega t}$. To first order in $U$ the wave equation becomes the Helmholtz
equation of a medium with a position-dependent refractive index,
\begin{equation}\label{eq:helmholtz}
\big[\nabla^2+\omega^2n^2(\bx)\big]\tilde\varphi=0,\qquad n^2=1-4U\, ,
\end{equation}
where $\nabla^2$ is the flat Laplacian. Physically, the lens acts like a lens of glass with index
$n=1-2U>1$: the wave is slowed down near the mass (the Shapiro delay) and bent towards it.  
Solutions of Eq.~\eqref{eq:helmholtz} are usually written in terms of the {\it amplification factor}
\begin{equation}\label{eq:Fdef}
F(\bx)\equiv\frac{\tilde\varphi(\bx)}{\tilde\varphi_0(\bx)} .
\end{equation}
Because the field equation is linear and time-independent, the complex amplification factor multiplies the unlensed signal in the frequency domain $\tilde h_{\rm lensed}(f)=F(f)\,\tilde h(f)$, modifying both its amplitude, through $F(f)$, and its phase, through $\arg F$.  
Since gravitational-wave interferometers measure the strain $h(t)$, both effects are observable, including relative image phases (e.g., the Morse phase of a saddle image) and the frequency dependence of $F(f)$ \cite{Dai2017,Ezquiaga2021}. By contrast, optical observations typically measure only $|F|^2$, as interference is washed out by the finite source size; gravitational-wave sources are effectively point-like, preserving phase information \cite{Leung2023}.

Assuming a point source sitting at the origin and using spherical coordinates $(r,\theta,\phi)$ centred on it, with $\tilde\varphi=F\,\e^{\ii\omega r}/r$, Eq.~\eqref{eq:helmholtz} gives \cite{NakamuraDeguchi1999,TakahashiNakamura2003}
\begin{equation}\label{eq:Feq}
\frac{\partial^2F}{\partial r^2}+2\ii\omega\,\frac{\partial F}{\partial r}+\frac{1}{r^2}\nabla_\theta^2F
=4\omega^2U\,F\, ,
\end{equation}
with $\nabla^2_\theta$ the Laplacian on the unit sphere. This is a second-order equation in $r$, the
distance from the source. One can reduce it to first order by dropping the
second derivative,
\begin{equation}\label{eq:ND_eikonal}
\Big|\frac{\partial^2F}{\partial r^2}\Big|\ll\Big|2\ii\omega\,\frac{\partial F}{\partial r}\Big| ,
\end{equation}
assuming that $\omega/|\partial\ln F/\partial r|\sim$ (scale on which $F$ varies)/(wavelength) $\gg1$. This approximation is
called the {\it eikonal approximation}\footnote{Apart from the condition adopted in the main text, ``eikonal'' has two inequivalent meanings in the literature.
In the geometric-optics (WKB) sense, the eikonal is the phase $S$ of $\phi=A\,e^{i\omega S}$, which solves $g^{\mu\nu}\partial_\mu S\,\partial_\nu S=0$ along null rays. The eikonal approximation is the leading order in $1/\omega$ and contains no diffraction; ``beyond-eikonal'' terms are the higher orders in $1/\omega$ along a ray~\cite{Bruyere:2026gnt}. In wave-optics lensing the name is also used for the stationary-phase sum over interfering images~\cite{Bonga:2024orc}.
In scattering theory, the eikonal limit is the small-angle regime $\ell\simeq\omega b\gg1$. The amplitude is the impact-parameter transform of $e^{i\delta(b)}-1$, where $\delta$ is the interaction integrated along the straight unperturbed trajectory~\cite{CarrilloGonzalez:2025gqm}.
}\cite{NakamuraDeguchi1999, Takahashi2004, Braga2024}.

Physically, it states that the envelope $F$ changes
little over one wavelength along the direction of propagation; Sec.~\ref{sec:eikonal_exact} shows that it
also selects the waves moving \emph{away} from the source and discards those moving back. 
In addition, it is standardly assumed that the waves stay close to the axis, $\theta\ll1$, so that the sphere can be replaced by its
tangent plane, $\bm\theta=(\theta\cos\phi,\theta\sin\phi)$, and $\nabla^2_\theta\to\nabla^2_{\bm\theta}$
(the \emph{paraxial} approximation). 
With these assumptions, Equation~\eqref{eq:Feq} becomes
\begin{equation}\label{eq:schrodinger_r}
\ii\,\frac{\partial F}{\partial r}=-\frac{1}{2\omega r^2}\nabla^2_{\bm\theta}F+2\omega U\,F .
\end{equation}
This is a Schr\"odinger equation in which the distance $r$ plays the role of time, the frequency $\omega$ the
role of the mass (with $1/\omega$ in place of $\hbar$), and $2\omega U$ the role of the potential. Its solution
is a Feynman path integral over the transverse trajectories $\bm\theta(r)$ of the ``particle'',
\begin{equation}\label{eq:pathint_r}
F(\bx_O)=\int\mathcal D\bm\theta(r)\;\exp\Big\{\ii\omega\int_0^{r_O}\dd r\Big[\frac{r^2}{2}\Big|\frac{\dd\bm\theta}{\dd r}\Big|^2-2U\big(r,\bm\theta\big)\Big]\Big\}\,.
\end{equation}

If the lens is thin compared with the distances $D_L$ (observer--lens), $D_{LS}$ (lens--source) and $D_S$
(observer--source), its potential can be projected onto a plane, $2U\to\delta(r-D_{LS})\,\hat\psi$, and between
the source and the lens plane, and between the lens plane and the observer, the paths are free. The path integral collapses to an ordinary integral over
the point where the path crosses the lens plane. In angular variables, with $\bm\theta$
the position of the crossing point seen by the observer and $\bm\beta$ the unlensed position of the source,
the result is
\begin{equation}\label{eq:diffint_phys}
F(\omega,\bm\beta)=\frac{\omega}{2\pi\ii}\,\frac{(1+z_L)D_LD_S}{cD_{LS}}\int\dd^2\theta\;\e^{\ii\omega t_d(\bm\theta,\bm\beta)},
\qquad
t_d=\frac{(1+z_L)D_LD_S}{cD_{LS}}\Big[\frac12|\bm\theta-\bm\beta|^2-\psi(\bm\theta)\Big],
\end{equation}
where the distances are angular-diameter distances, $z_L$ is the redshift of the lens and $\psi$ is the
lensing potential, the projected Newtonian potential in angular units, whose gradient is the deflection angle.
The function $t_d$ is the arrival time of the broken path through $\bm\theta$, relative to the unlensed path:
its first term is the geometric delay of the detour, its second the Shapiro delay.

For $w\to\infty$ the integrand oscillates rapidly except near the points where $\Tf$ is
stationary,
\begin{equation}\label{eq:lenseq}
\nabla_{\bx}\Tf(\bx,\by)=0\quad\Longleftrightarrow\quad \by=\bx-\nabla_{\bx}\psi(\bx)\qquad\text{(lens equation)},
\end{equation}
which is Fermat's principle: the rays are the paths of stationary arrival time. This is the standard result of Geometric Optics \cite{Isaacson:1967sln,Isaacson1968GravitationalRI}.  Expanding $\Tf$ to second
order around each solution $\bx_j$ and doing the Gaussian integral gives
\begin{equation}\label{eq:FGO_general}
F_{\rm GO}=\sum_j|\mu_j|^{1/2}\exp\big(\ii w\Tf_j-\ii\pi n_j\big),\qquad
\mu_j=\Big[\det\frac{\partial\by}{\partial\bx}\Big]^{-1}_{\bx_j},
\end{equation}
with $n_j=0$, $\frac12$ and $1$ for a minimum, a saddle and a maximum of $\Tf$ (the Morse index divided by two).
Each image contributes with its arrival time $t_j=4M\Tf_j$, an amplitude set by the curvature of the arrival
time (flux conservation along a bundle of rays) and a phase jump each time the bundle has passed through a
focus. The magnifications do not depend on frequency: geometric optics is achromatic. 
\subsection{The point mass}\label{sec:pointmass}
For a point mass, Eq.~\eqref{eq:helmholtz} becomes a Coulomb problem,
\begin{equation}\label{eq:coulomb}
\Big(\nabla^2+\omega^2+\frac{\kappa}{r}\Big)\tilde\varphi=0,\qquad \kappa\equiv4M\omega^2=w\,\omega,
\end{equation}
equivalent to an attractive Coulomb potential with energy-dependent coupling. We define the Sommerfeld parameter $\nu\equiv w/2=2M\omega$ and
the lens potential is $\psi=\theta_E^2\ln|\bm\theta|$, where $\theta_E^2=4GMD_{LS}/(c^2D_LD_S)$. Rescaling $\bm\theta=\theta_E\bx$ and $\bm\beta=\theta_E\by$ gives $t_d=(4GM_z/c^3)\,\Tf(\bx,\by)$ and the dimensionless diffraction integral
\begin{equation}\label{eq:kirchhoff}
F(w,\by)=\frac{w}{2\pi\ii}\int\dd^2x\;\e^{\ii w\Tf(\bx,\by)},\qquad
\Tf(\bx,\by)=\frac12|\bx-\by|^2-\ln|\bx|\,.
\end{equation}
The prefactor is a normalization that ensures $F=1$, while the physical lens-plane radius is $\xi=x\xi_E$, with $\xi_E=\theta_ED_L=(4GMD_{\rm eff})^{1/2}/c$ and $D_{\rm eff}=D_LD_{LS}/D_S$. Equation~\eqref{eq:kirchhoff} is the Kirchhoff (or Fresnel--Kirchhoff) diffraction integral for a thin phase screen \cite{SEF1992,FPT2019}. The diffraction integral can also be evaluated exactly, yielding a confluent hypergeometric function \cite{Peters1974,DeguchiWatson1986,Nakamura1998}:
\begin{equation}\label{eq:Fwy}
F(w,y)=\Big(\!-\frac{\ii w}{2}\Big)^{\ii w/2}\Gamma\Big(1-\frac{\ii w}{2}\Big)\,
\hF\Big(\frac{\ii w}{2},1;\frac{\ii w}{2}y^2\Big)\,.
\end{equation}
It is important to notice that in the limit of long wavelengths $w\to0$ one recovers $F\to1$ as  waves with wavelengths much larger than the Schwarzschild radius are essentially unaffected by the lens \cite{Leung2023}.

\begin{figure}[h!]\centering
\includegraphics[width=\textwidth]{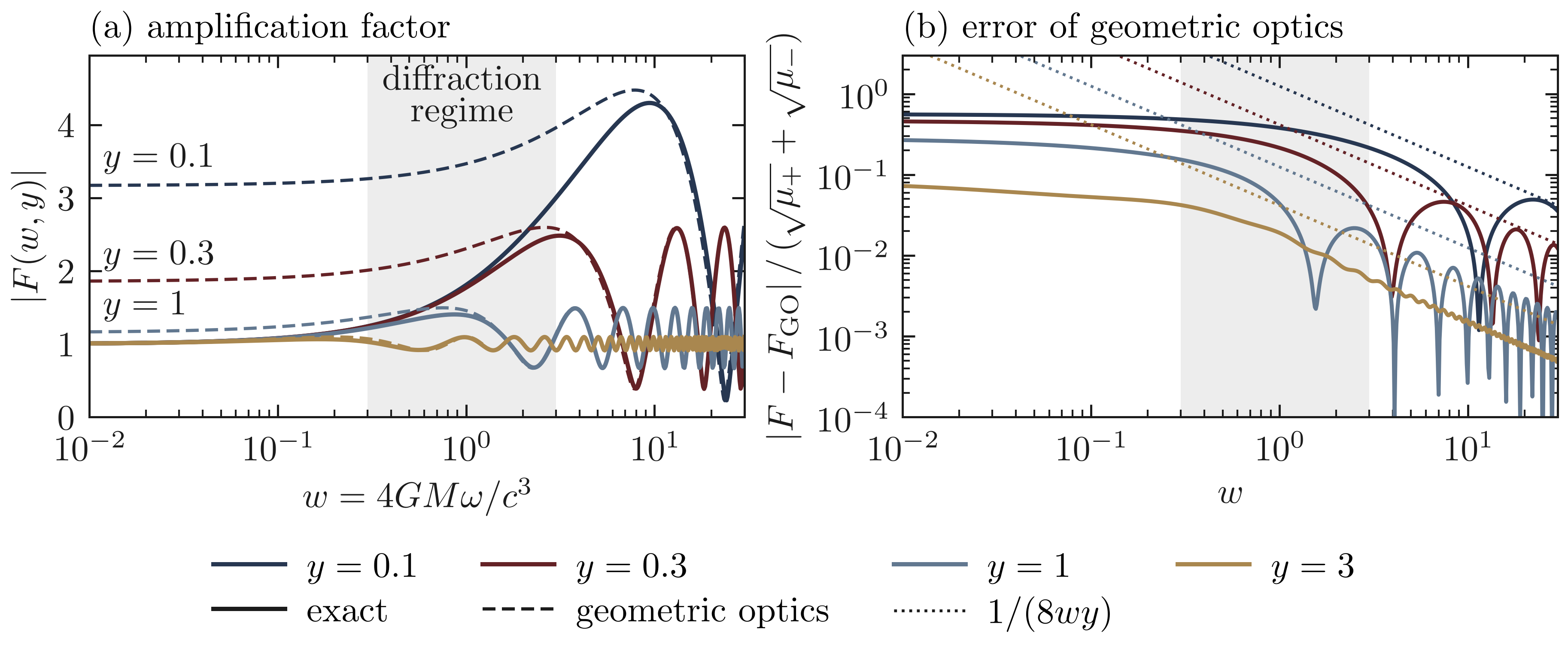}
\caption{Point-lens amplification factor. (a) $|F(w,y)|$ from Eq.~\eqref{eq:Fwy} (solid) and from geometric
optics, Eq.~\eqref{eq:FGO} (dashed). For $w\ll1$ the exact $F$ tends to 1, while geometric optics keeps
predicting the full magnification. (b) Error of geometric optics, normalized by $\sqrt{\mu_+}+\sqrt{\mu_-}$; the
dotted lines are $1/(8wy)$. The grey band marks the diffraction regime $0.3\le w\le3$. }\label{fig:Fwy}
\end{figure}

When studying the GO limit, one finds that the lens equation has two images on opposite sides of the lens, with radii and signed magnifications
\begin{equation}\label{eq:images}
x_\pm=\tfrac12\big(\sqrt{y^2+4}\pm y\big),\qquad
\mu_\pm=\frac{y^2+2}{2y\sqrt{y^2+4}}\pm\frac12 ,
\end{equation}
where $x_+$ is a minimum of $\Tf$ and $x_-$ a saddle \cite{Refsdal1964}. Their time-delay difference is
\begin{equation}\label{eq:delay}
\Delta\tauh=\Tf_--\Tf_+=\tfrac12y\sqrt{y^2+4}+\ln\frac{\sqrt{y^2+4}+y}{\sqrt{y^2+4}-y}\simeq2y\quad(y\ll1),
\end{equation}
in units of $4M$, giving $\Delta t\simeq8My$ for small offsets. The amplification is
\begin{equation}\label{eq:FGO}
F_{\rm GO}=\sqrt{\mu_+}\,\e^{\ii w\Tf_+}-\ii\sqrt{\mu_-}\,\e^{\ii w\Tf_-},
\end{equation}
where $-\ii=\e^{-\ii\pi/2}$ is the saddle's Morse phase. Notice that the full geometric-optics regime requires both $w\gg1$ and $wy\gg1$. A quick way to see that high frequency alone is
not enough is the on-axis value: Eq.~\eqref{eq:Fwy} gives $|F(w,0)|^2=\pi w/(1-\e^{-\pi w})$, finite at every $w$,
whereas geometric optics predicts $(\sqrt{\mu_+}+\sqrt{\mu_-})^2\simeq2/y$, which diverges on the axis; the two can
agree only for $wy\gtrsim1$. Figure~\ref{fig:Fwy}(b) shows the error of Eq.~\eqref{eq:FGO}, which scales as
$\approx1/(8wy)$ at small $y$~\cite{Takahashi2004}.

\begin{figure}[h!]\centering
\includegraphics[width=\textwidth]{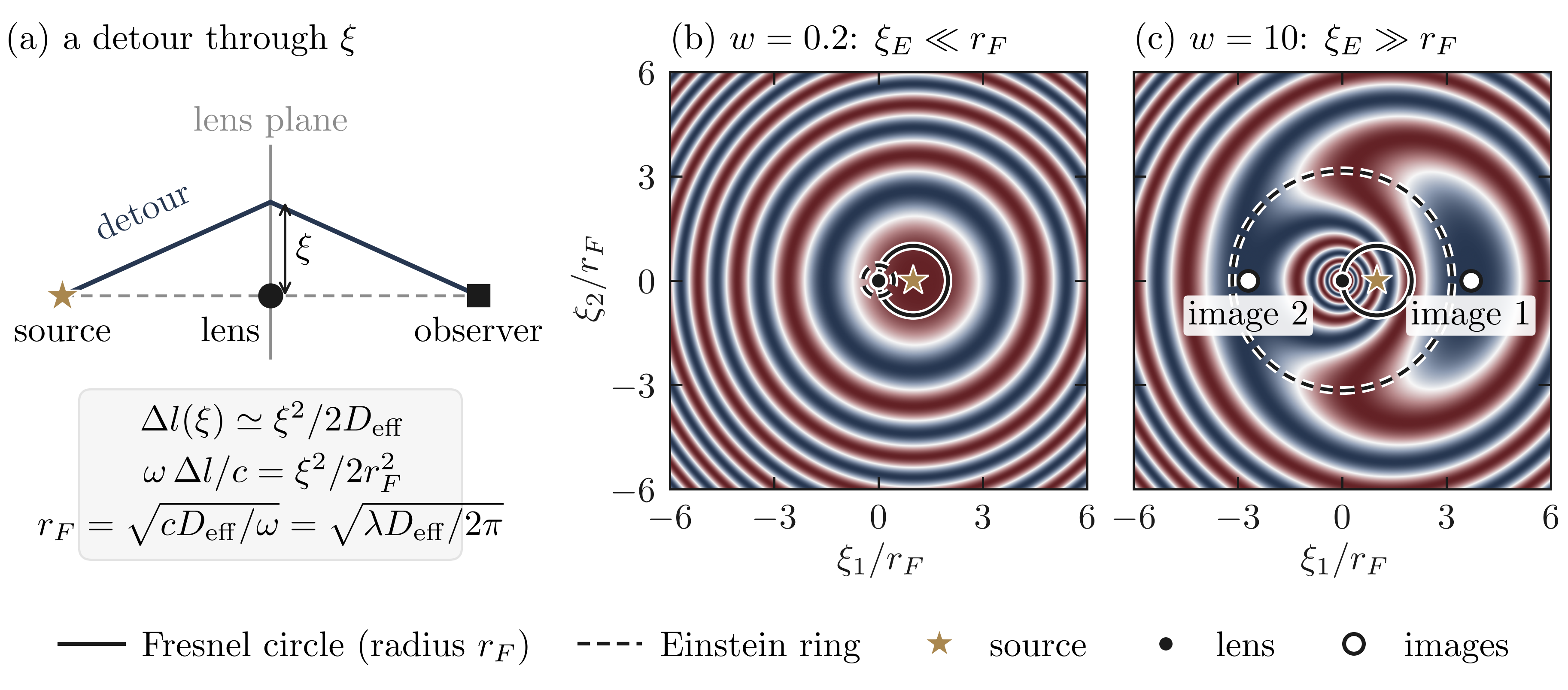}
\caption{The Fresnel radius. (a) Each lens-plane point defines a detour path, whose extra phase is one radian at $\xi=r_F$. (b, c)
The lens plane in units of $r_F$: colour is the cosine of the phase (detour plus Shapiro) of each point, for a source one Fresnel
radius off axis (star), with the Fresnel circle (solid) and the Einstein ring (dashed). For $w=0.2$ the lens barely distorts the
central zone; for $w=10$ the zones cancel except near the two images.}\label{fig:fresnel}
\end{figure}

Both limits have a simple meaning in terms of the Fresnel scale. A path through a lens-plane point at distance $\xi$ from
the unlensed one is longer by $\xi^2/2D_{\rm eff}$, so lens-plane points add in phase only within the Fresnel radius
$r_F=(cD_{\rm eff}/\omega)^{1/2}$ (Fig.~\ref{fig:fresnel}), and $w=(\xi_E/r_F)^2$: for $w\ll1$ the Einstein ring lies inside
the first Fresnel zone and the lens is barely felt, $F\simeq1$, while for $w\gg1$ it spans many zones, which cancel
everywhere except near the images~\cite{Jow2023}. Finally, since $F$ multiplies the spectrum, in the time domain the
lensed signal is the unlensed one convolved with the impulse response of the lens, which is fixed by the delay density
\begin{equation}\label{eq:rho_std}
\rh(\tauh;\by)=\int\dd^2x\;\delta\big(\tauh-\Tf(\bx,\by)\big)\,,
\end{equation}
the area of the lens plane per unit Fermat delay~\cite{UlmerGoodman1995,Tambalo2023,GLoW2025}; its singularities are the
images, a step at a minimum and a logarithm at a saddle.

\section{The proper-time (worldline) technique}\label{sec:propertime}

We saw that the diffraction integral assumes a scalar wave propagating through a weak, static, thin lens, with source and observer in the far field, under the eikonal and paraxial approximations. These two latter approximations are used in order to obtain a Schr\"odinger-like equation, which solution can be expressed in terms of a particle path integral Eq.~\eqref{eq:pathint_r}. 

In order to obtain a first-order evolution equation from Eq~\eqref{eq:helmholtz} without having to rely on any approximation, one can introduce an auxiliary evolution parameter, the {\it proper time}, that is not a coordinate at all. This is Fock's and Schwinger's construction \cite{Fock1937,Nambu1950,Feynman1950,Schwinger1951}, the basis of the
worldline formalism of quantum field theory \cite{Strassler1992,Schubert2001,Bastianelli:2013tsa,
EdwardsSchubert2019}, and it is exact. It was first brought it into GW lensing in \cite{Braga2024,Carrillo-Gonzalez:2026oov}.
The price of this approach is an extra integral, over the proper time, which carries much of the physics.
Its reward is that every approximation is taken a posteriori, as a saddle point of an exact integral,
so its validity can be judged.
In this section we do a short review of this formalism applied to the Coulomb problem. We keep the frequency explicit as the inverse of $\hbar$ throughout, so that we remain under the umbrella of the Hamilton analogy \cite{Masoliver_2009,goldstein2002classical}.

Introducing a point source, Eq.~\eqref{eq:coulomb} can be written as $\hat H_\omega G_\omega=\delta^3(\bx-\bx')$, with
$\hat H_\omega=-\nabla^2-\omega^2n^2$. We measure the momentum in units of the frequency and factor $\omega^2$ out,
\begin{equation}\label{eq:Homega}
\hat H_\omega=\omega^2\hat H,\qquad \hat H=\hat\bp^2-1+V(\bx),\qquad \hat\bp=-\frac{\ii}{\omega}\nabla,\qquad V=1-n^2=-\frac{4M}{r}.
\end{equation}
The momentum operator is $\hat\bp=-\frac{\ii}{\omega}\nabla$, while the potential $V$ and the symbol
$H(\bx,\bp)=\bp^2-n^2(\bx)$ of $\hat H$ do not depend on the frequency. The Schwinger representation $(a-\ii\epsilon)^{-1}=\ii\omega\int_0^\infty\dd\tau\,\e^{-\ii\omega\tau(a-\ii\epsilon)}$,
valid for real $a$, $\epsilon>0$ and any $\omega>0$ \cite{Schwartz:2014sze}, applied to $\hat H=\hat H_\omega/\omega^2$ gives
\begin{equation}\label{eq:schwinger}
G_\omega(\bx,\bx')=\frac{\ii}{\omega}\int_0^\infty\dd\tau\;\e^{\ii\omega\tau(1+\ii\epsilon)}\,\tilde G_\omega(\bx,\bx';\tau),\qquad
\tilde G_\omega(\bx,\bx';\tau)=\big\langle\bx\big|\e^{-\ii\omega\tau(\hat\bp^2+V)}\big|\bx'\big\rangle\, .
\end{equation}
The proper time $\tau$ has the dimension of a length. It is related to the more common Schwinger parameter $s$, conjugate to $\omega^2$
in $\e^{\ii s\omega^2}$, by $\tau=\omega s$, which is the variable used in Ref.~\cite{Braga2024}. The kernel $\tilde G_\omega$ satisfies
\begin{equation}\label{eq:schrod_s}
\frac{\ii}{\omega}\,\frac{\partial \tilde G_\omega}{\partial \tau}=\Big(-\frac{1}{\omega^2}\nabla^2+V\Big)\tilde G_\omega,\qquad \tilde G_\omega(\tau=0)=\delta^3(\bx-\bx')\,,
\end{equation}
a Schrödinger equation in the auxiliary time $\tau$ with $\hbar=1/\omega$, describing a fictitious particle of mass $1/2$ in the
frequency-independent potential $V$. The prescription $1\to1+\ii\epsilon$ (i.e.\ $\omega^2\to\omega^2+\ii\epsilon$) enforces outgoing boundary conditions.

Time-slicing the evolution operator yields the phase-space path integral \cite{Feynman1965}
\begin{equation}\label{eq:tildeG}
\tilde G_\omega=\int\mathcal D\bx\,\mathcal D\bp\,\exp\left\{\ii\omega\int_0^\tau\dd\tau'\,[\bp\cdot\dot\bx-(\bp^2+V)]\right\},
\end{equation}
with $\dot\bx=\dd\bx/\dd\tau'$ and $\tau'\in[0,\tau]$ the running proper time. Together with the integration over $\tau$, this is equivalent to the first-order worldline action \cite{1976PhLB...64..435B,Brink:1976zz,Schubert2001,BastianelliVN2006}
\begin{equation}\label{eq:einbein}
S[\bx,\bp,e]=\omega\int_0^1\dd\varsigma\Big[\bp\cdot\frac{\dd\bx}{\dd\varsigma}-e(\varsigma)\,H(\bx,\bp)\Big],
\qquad H=\bp^2-1+V=\bp^2-n^2(\bx) .
\end{equation}

Although gravitational waves propagate along null worldlines, whose physical proper time vanishes, the parameter $\tau$ has a geometric interpretation as the worldline length defined by an auxiliary metric, encoded by the \emph{einbein} $e(\varsigma)$. The action is invariant under reparametrizations $\varsigma\to f(\varsigma)$ provided $e\,\dd\varsigma$ is unchanged, so $\varsigma$ is an arbitrary worldline parameter. Varying the einbein imposes $H=0$, which for a static lens yields $|\bp|=n$ and hence the null-propagation condition $|\omega\bp|^2=\omega^2n^2$. Nevertheless, the invariant length $\tau=\int_0^1e\,\dd\varsigma$ must still be integrated over, reproducing the Schwinger integral in Eq.~\eqref{eq:schwinger} \cite{Guralnik2018,CMNP1986,Polyakov1987,Teitelboim1982,HenneauxTeitelboim1992}. The gauge choice also determines the worldline clock. For the frequency-independent action $S/\omega$, Hamilton's equations are
\begin{equation}\label{eq:hamilton}
\frac{\dd\bx}{\dd\tau'}=2\bp,\qquad \frac{\dd\bp}{\dd\tau'}=-\nabla V=-\frac{4M\,\bx}{r^3}.
\end{equation}
On shell, these coincide with the ray equations in a medium of index $n$. Introducing the affine parameter $\lambda$ through $\dd\bx/\dd\lambda=\bp$, for which $|\dd\bx/\dd\lambda|=n$, gives $\tau'=\lambda/2$. Thus, along a classical worldline, the proper-time parameter is half the affine parameter of the ray.
If both endpoint
\emph{times} were fixed, the total time could not vary and Fermat's principle could not emerge; for a
monochromatic wave one must fix the \emph{energy} instead, which requires a boundary term. Integrating out the
einbein and the momenta at their saddle then gives the reduced action
$S_r=\omega\int|\dd\bx|\,n(\bx)$, the optical length: Fermat's principle is the proper-time saddle point of the
worldline action \cite{FPT2019}. The same result can be obtained by
requiring stationary in the proper time only and assuming
both the paraxial and thin-lens approximations \cite{Braga2024}, the paths become straight lines broken at one point of the
lens plane, the path integral collapses into the
diffraction integral~\eqref{eq:kirchhoff} .

Requiring in addition that
the action is stationary under variations of the path gives the classical rays.
\section{An exact solution for a point mass: the Duru--Kleinert gauge}\label{sec:DK}

For a point mass $V=-4M/r$, and the Schwinger representation~\eqref{eq:schwinger}, with its path integral written
out, reads
\begin{equation}\label{eq:GS_pi}
G_\omega(\bx,\bx')=\frac{\ii}{\omega}\int_0^\infty\dd\tau\;\e^{\ii\omega\tau(1+\ii\epsilon)}
\int_{\bx(0)=\bx'}^{\bx(\tau)=\bx}\mathcal D\bx\;
\exp\Big\{\ii\omega\int_0^\tau\dd\tau'\Big[\frac14\Big|\frac{\dd\bx}{\dd\tau'}\Big|^2+\frac{4M}{r}\Big]\Big\}.
\end{equation}
The frequency multiplies the whole exponent, and the lens sits inside the path integral as the frequency-independent
Coulomb potential $4M/r$. At fixed $\tau$ the path integral is not Gaussian, and keeping only its classical path is the WKB
approximation, the leading order in $1/\omega$. Duru and
Kleinert \cite{DuruKleinert1979,DuruKleinert1982} solved the path integral of the hydrogen atom, which has the same
structure as Eq.~\eqref{eq:GS_pi}, with two changes: a change of the clock along the worldline, and a change of variables from
$\mathbb R^3$ to $\mathbb R^4$. 

\begin{figure}
    \centering
    \includegraphics[width=0.8\linewidth]{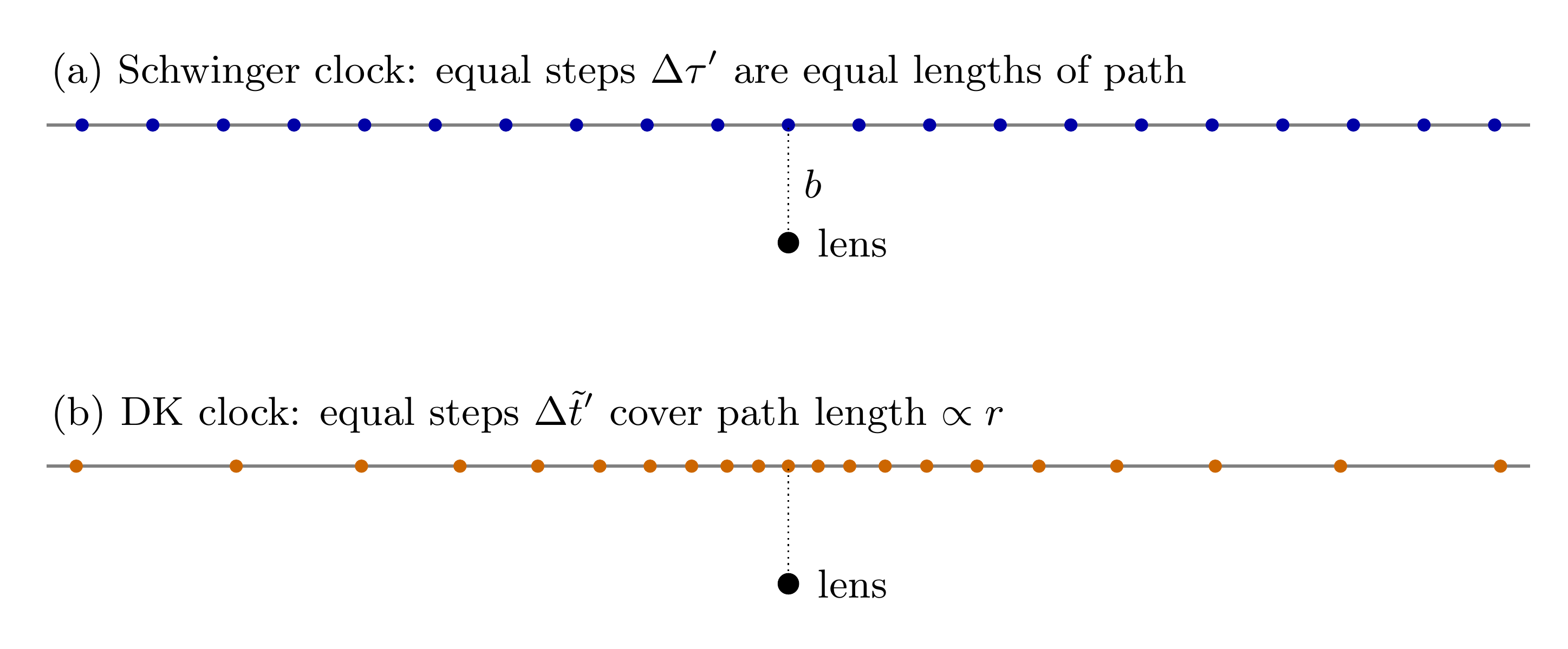}
    \caption{The same straight path, at impact parameter b, marked at equal intervals of the two clocks. With $d
\tilde t'= d\tau'/r$ the DK ticks crowd near the lens; their positions are $x_k = b\sinh(k\delta)$. A path that passes closer to the lens collects more DK time, i.e. more Shapiro delay.}
    \label{fig:DK_vs_schwinger}
\end{figure}

The clock is fixed by the gauge of the einbein in the action~\eqref{eq:einbein}.  The Duru--Kleinert (DK) gauge makes the einbein proportional to the distance from the lens \cite{Kleinert2009,DuruKleinert1979,DuruKleinert1982},
\begin{equation}\label{eq:DKgauge}
e(\varsigma)=\tilde t\,r\big(\bx(\varsigma)\big),\qquad \tilde t=\int_0^1\dd\varsigma\,\frac{e(\varsigma)}{r\big(\bx(\varsigma)\big)},
\end{equation}
and its invariant modulus is $\tilde t$, the DK time, which is dimensionless because $e$ has the dimension of a length. Along a path the Schwinger clock advances by $\dd\tau'=e\,\dd\varsigma$
and the DK clock by $\dd\tilde t'=e\,\dd\varsigma/r$, so that
\begin{equation}\label{eq:DKclock}
\dd\tilde t'=\frac{\dd\tau'}{r}.
\end{equation}
As shown in Fig.~\ref{fig:DK_vs_schwinger}, the DK clock runs faster near the lens and slower farther away. It also has a direct physical interpretation: along a classical worldline, $\dd\tau'=\dd\lambda/2$ and $\dd l=|\bp|\,\dd\lambda\simeq\dd\lambda$, so $\tilde t=\int\dd\tau'/r=\int\dd l/(2r)$. Since the Shapiro delay in the metric~\eqref{eq:metric} is $\Delta t_{\rm Sh}=2M\int\dd l/r$, we obtain
\begin{equation}\label{eq:kappaS}
4M\tilde t=4M\int\frac{\dd l}{2r}=\Delta t_{\rm Sh},\qquad w\tilde t=\omega\,\Delta t_{\rm Sh}.
\end{equation}
Thus, $\tilde t$ measures the Shapiro delay in units of $4M$, and $\e^{\ii w\tilde t}$ in Eq.~\eqref{eq:GDK3} is the corresponding Shapiro phase.
In this gauge the term $eH$ of the
action becomes $\tilde t\,rH$, with
\begin{equation}\label{eq:rH}
rH=r\bp^2-r-4M .
\end{equation}
Multiplying the constraint by $r>0$ does not change where it vanishes, so the classical worldlines are the same in
both gauges. What changes is the role of the two constants of $H=p^2-1-4M/r$: the vacuum term $1$, which comes from
$\omega^2$ in the Helmholtz equation and is the energy of the free wave, and the lens strength $4M$. Particularly, 
the lens strength becomes a constant term, playing the part of an energy, while the vacuum term now multiplies $r$ and,
after the Kustaanheimo--Stiefel map below, becomes the strength of an inverted oscillator potential. At the operator level, the statement that the two
gauges describe the same wave is the exact identity
\begin{equation}\label{eq:DKop}
\hat H^{-1}=\big(r\hat H\big)^{-1}r ,
\end{equation}
and applying Schwinger's formula, $(a-\ii\epsilon)^{-1}=\ii\omega\int_0^\infty\dd\tilde t\,\e^{-\ii\omega\tilde t(a-\ii\epsilon)}$, to the operator $r\hat H=r\hat\bp^2-(1+\ii\epsilon)r-4M$ instead of
$\hat H$ gives
\begin{equation}\label{eq:GDK3}
G_\omega(\bx,\bx')=\frac{\ii}{\omega}\,r'\int_0^\infty\dd\tilde t\;\e^{\ii w\tilde t}\,
\int_{\bx(0)=\bx'}^{\bx(\tilde t)=\bx}\mathcal D\bx\;
\exp\Big\{\ii\omega\int_0^{\tilde t}\dd\tilde t'\Big[\frac{1}{4r}\Big|\frac{\dd\bx}{\dd\tilde t'}\Big|^2+r\Big]\Big\}\,,
\end{equation}
where $ w=4M\omega$ and $r'=|\bx'|$ is the distance of the source from the lens: the rightmost $r$ in
Eq.~\eqref{eq:DKop} acts on the source point. Equivalently, the Green function of $r\hat H$ solves $\hat H g=\delta^3(\bx-\bx')/r'$, and the factor $r'$ restores a unit source.
 Comparing with
Eq.~\eqref{eq:GS_pi}, the weight of the proper-time integral has changed from $\e^{\ii\omega\tau}$ to
$\e^{\ii\omega\cdot4M\tilde t}$: the DK time is conjugate to the lens strength, and the whole lens now sits in this factor, while
the frequency still multiplies the whole exponent. Path by path the two exponents are the same number: with
$\dd\tau'=r\,\dd\tilde t'$, so that 
\begin{equation}\label{eq:sameaction}
\omega\Big\{\tau+\int_0^\tau\dd\tau'\Big[\frac14\Big|\frac{\dd\bx}{\dd\tau'}\Big|^2+\frac{4M}{r}\Big]\Big\}
=\omega\Big\{4M\tilde t+\int_0^{\tilde t}\dd\tilde t'\Big[\frac{1}{4r}\Big|\frac{\dd\bx}{\dd\tilde t'}\Big|^2+r\Big]\Big\}\,.
\end{equation}
We see directly that the Coulomb potential has disappeared from the paths. In its place there is the term $r$, and the kinetic term
has acquired the position-dependent mass $1/2r$. Equation~\eqref{eq:GDK3} is therefore still not Gaussian, and it is
only formal, because with a position-dependent mass the time-sliced path integral depends on how the slicing is
done \cite{Kleinert2009,GroscheSteiner1998}. To tackle these problems, we apply the Kustaanheimo--Stiefel (KS) map \cite{KS1965}. We write a point $\bu\in\mathbb R^4$ as a
complex doublet $z=(u_1+\ii u_2,\,u_3+\ii u_4)$ and define
\begin{equation}\label{eq:KS}
x_i=z^\dagger\sigma_iz\qquad(i=1,2,3),
\end{equation}
with $\sigma_i$ the Pauli matrices. Exploiting the completeness relation
$\sum_i(\sigma_i)_{ab}(\sigma_i)_{cd}=2\delta_{ad}\delta_{bc}-\delta_{ab}\delta_{cd}$, one finds
\begin{equation}\label{eq:KSr}
r=|\bx|=z^\dagger z=|\bu|^2\,,
\end{equation}
and expanding the Hermitian matrix $zz^\dagger$ in the basis $\{\mathbb 1,\sigma_i\}$ gives $zz^\dagger=\frac12(r+\bx\cdot\bm\sigma)$.
The point $\bx$ therefore fixes the projector $zz^\dagger$, and with it $z$ up to a phase: all the points
$\e^{\ii\theta}z$ of a circle in $\mathbb R^4$ are mapped to the same $\bx$. This circle is the fibre over $\bx$, with
$\theta$ being the fibre angle. 
The operator of Eq.~\eqref{eq:GDK3}, acting on fibre-independent
functions, is
\begin{equation}\label{eq:hosc}
r\hat H=-\frac{r}{\omega^2}\nabla_x^2-r-4M=-\frac{1}{4\omega^2}\nabla_u^2-|\bu|^2-4M\equiv\hat h-4M,
\qquad \hat h=\tfrac14\hat\bp_u^2-|\bu|^2,\quad \hat\bp_u=-\frac{\ii}{\omega}\nabla_u .
\end{equation}
This is a four-dimensional isotropic harmonic oscillator with a negative spring constant, an inverted oscillator,
appropriate for unbound, scattering orbits, with $\hbar=1/\omega$. Its energy is fixed to the lens strength $4M$. Because
Eq.~\eqref{eq:hosc} is an operator identity, no term of higher order in $1/\omega$ appears. The same simplification appears in the paths of Eq.~\eqref{eq:GDK3}. Using Eq.~\eqref{eq:KS}, $\dd\bx/\dd\tilde t'=2\,{\rm Re}(z^\dagger\bm\sigma\dot z)$, where the dot denotes $\dd/\dd\tilde t'$. The completeness relation then yields
\begin{equation}\label{eq:xdotKS}
\Big|\frac{\dd\bx}{\dd\tilde t'}\Big|^2=4r|\dot\bu|^2-J^2,\qquad J=2\,{\rm Im}(z^\dagger\dot z)=2\big(u_1\dot u_2-u_2\dot u_1+u_3\dot u_4-u_4\dot u_3\big),
\end{equation}
where $J$ generates fibre rotations $z\to\e^{\ii\alpha}z$ and is expressed in terms of the momentum $\bp_u=2\dot\bu$ conjugate to $\bu$. For paths with no fibre motion, $J=0$, the kinetic term in Eq.~\eqref{eq:GDK3} reduces to $|\dd\bx/\dd\tilde t'|^2/4r=|\dot\bu|^2$. Since $r=|\bu|^2$, the Lagrangian becomes $|\dot\bu|^2+|\bu|^2$: a quadratic form with constant mass. The fibre integral then projects the four-dimensional problem back onto the physical three-dimensional space.

We rewrite the Green function in the Schwinger representation in the DK time,
\begin{equation}\label{eq:gS}
g(\bu,\bu')=\frac{\ii}{\omega}\int_0^\infty\dd\tilde t\;\e^{\ii w\tilde t}\,\langle\bu|\e^{-\ii\omega\tilde t\hat h}|\bu'\rangle ,
\end{equation}
 and writing the kernel of $\hat h$ as a
path integral, we obtain the Green function in the DK gauge:
\begin{equation}\label{eq:GDK4}
G_\omega(\bx,\bx')=\frac{\ii}{8\omega}\int_0^{2\pi}\dd\theta'\int_0^\infty\dd\tilde t\;\e^{\ii w\tilde t}
\int_{\bu(0)=\e^{\ii\theta'}\bu'}^{\bu(\tilde t)=\bu}\mathcal D\bu\;
\exp\Big\{\ii\omega\int_0^{\tilde t}\dd\tilde t'\,\Big[\Big|\frac{\dd\bu}{\dd\tilde t'}\Big|^2+(1+\ii\epsilon)|\bu|^2\Big]\Big\}.
\end{equation}
This form can be compared directly with the Schwinger representation~\eqref{eq:GS_pi}: in both, the frequency multiplies the entire exponent, while the remaining terms are frequency-independent. The path integral in Eq.~\eqref{eq:GDK4} describes a free inverted oscillator and can be evaluated exactly, yielding the Mehler kernel
\begin{equation}\label{eq:mehler}
\int_{\bu(0)=\bu''}^{\bu(\tilde t)=\bu}\mathcal D\bu\;\e^{\ii\omega\int_0^{\tilde t}\dd\tilde t'(\dot\bu^2+\bu^2)}
=-\frac{\omega^2}{\pi^2\sinh^2\tilde t}\;\e^{\ii\omega\Acl(\tilde t,\theta')}\,,
\end{equation}
with 
\begin{equation}\label{eq:Acl}
\Acl(\tilde t,\theta')=(r+r')\coth\tilde t-\frac{2\,\bu\cdot\bu''}{\sinh\tilde t}.
\end{equation}
Then, the remaining integral over the initial fibre angle can then be recast in a convenient form. Details are given in Appendix~\ref{App:A}. The exact propagator now takes the form
\begin{equation}\label{eq:GDK}
G_\omega(\bx,\bx')=-\frac{\ii\omega}{4\pi}\int_0^\infty\frac{\dd\tilde t}{\sinh^2\tilde t}\;\e^{\ii w\tilde t}\;
\e^{\ii\omega(r+r')\coth\tilde t}\;J_0\Big(\frac{2\omega\varpi}{\sinh\tilde t}\Big)\,,
\end{equation}
with $J_0(q)=\frac1{2\pi}\int_0^{2\pi}\dd\theta\,\e^{\ii q\cos\theta}$ and $\varpi$ defined in Eq.~\eqref{eq:varpi}.
It is exact within the linearized field, for all frequencies, positions and angles.
\subsection{Analyzing the stationary contributions}\label{sec:constraints}
After performing the Gaussian integral, Eq~\eqref{eq:GDK4} remains with the integrals over $\tilde t$ and $\theta'$, with
phase
\begin{equation}\label{eq:Psi_SL}
\Psi(\tilde t,\theta')=\omega\big[4M\tilde t+\Acl(\tilde t,\theta')\big],
\end{equation}
where $\Acl$, the action of the classical oscillator orbit.
Since $\omega$ multiplies the whole phase, at high frequency the integrals in $\{\tilde t,\theta'\}$ can be
evaluated by stationary phase. By Hamilton--Jacobi theory, the
derivatives of the classical action with respect to the duration of the orbit and to a rotation of its initial point along the
fibre are conserved quantities of the orbit,
\begin{equation}\label{eq:HJ}
\frac{\partial\Acl}{\partial\tilde t}=-\Eosc,\qquad \frac{\partial\Acl}{\partial\theta'}=\pm J,
\end{equation}
where $\Eosc=\dot\bu^2-\bu^2$ is the energy of the oscillator orbit and $J$ its fibre charge, Eq.~\eqref{eq:xdotKS}. The two
stationarity conditions are therefore
\begin{equation}\label{eq:statS}
\partial_{\tilde t}\Psi=\omega\,(4M-\Eosc)=0 ,
\end{equation}
the Hamiltonian constraint in the DK gauge, $rH=0$, and
\begin{equation}\label{eq:stattheta}
\partial_{\theta'}\Psi=\pm\omega J=0 ,
\end{equation}
the KS constraint, which states that the orbit does not move along the fibre.

Each pair $(\tilde t,\theta')$ defines a trial orbit, which the KS map projects into three dimensions. Using
$\dd\tau'=r\,\dd\tilde t'$, Eq.~\eqref{eq:DKclock}, the projected momentum is
$\bp_x=\frac12\dd\bx/\dd\tau'=(\dd\bx/\dd\tilde t')/(2r)$, and combining Eq.~\eqref{eq:xdotKS} with $\dot\bu^2=\Eosc+r$
gives
\begin{equation}\label{eq:proj}
|\bp_x|^2=\frac{4r\dot\bu^2-J^2}{4r^2}=1+\frac{\Eosc}{r}-\frac{J^2}{4r^2}.
\end{equation}
The rays of geometric optics, i.e.\ the classical trajectories of the lens, have $|\bp_x|^2=1+4M/r$ and lie in the
plane of source, lens and observer. A trial orbit differs from them in two ways: for $\Eosc\neq4M$ it is deflected as if
the lens had a different strength, so that it passes the lens at a different distance and accumulates a different
Shapiro delay, and for $J\neq0$ it leaves the source--lens--observer plane, passing the lens at a different azimuth.
Every trial orbit therefore solves the equations of motion but violates the constraints, and stationarity in $\tilde t$
and $\theta'$ imposes $\Eosc=4M$ and $J=0$, selecting the two geometric-optics rays, which are planar Rutherford
orbits~\cite{Rutherford:1911zz,Pijnenburg:2022pug} and correspond to the two images. The other trial orbits are not
discarded: they are the paths through all the other points around the lens, and their interference produces the
wave-optics effects. Orbits with a DK time close to the stationary one give the finite radial width of the diffraction
pattern, while orbits with $J\neq0$ carry the wave around the lens at every azimuth; when $wy\lesssim1$ they add in
phase with the two rays, the wave reaches the observer from all around the lens, as in an Einstein ring, and the
magnification stays finite where geometric optics diverges (Sec.~\ref{sec:lensplane}).

This is the counterpart of the trial paths of the diffraction integral~\cite{CarrilloGonzalez:2025gqm,Braga2024}.
There every path is on the energy shell but is kinked at the lens plane, and stationarity in the lens-plane point
removes the kink; here every path solves the equations of motion, and stationarity in $\tilde t$ and $\theta'$
enforces the constraints. For an observer at infinity the two phase functions coincide (Sec.~\ref{sec:lensplane}), so
the two pictures give the same images and magnifications.

\subsection{Observer at infinity}\label{sec:lensplane}

We now place the observer at $\bx=r\bn$ with $r\to\infty$, keep the source at $\bx'$, and define
\begin{equation}\label{eq:eta}
\eta\equiv r'+\bn\cdot\bx',\qquad \zeta\equiv\omega\eta ,
\end{equation}
the extra length of the route source $\to$ lens $\to$ observer over the direct route and the same
in radians. Shifting the DK time by the large logarithm accumulated on the long leg, $\tilde t=\tdk+\frac12\ln(2\omega r)$, the
propagator~\eqref{eq:GDK} becomes
\begin{equation}\label{eq:far}
G_\omega(r\bn,\bx')\to\frac{\e^{\ii\omega r+\ii\nu\ln2\omega r}}{4\pi r}\,\e^{-\ii\omega\bn\cdot\bx'}\,F_\omega(\bx'),
\qquad
\e^{-\ii\zeta}F_\omega=-\frac{\ii}{\pi}\int_{-\infty}^\infty\!\dd\tdk\int_0^{2\pi}\!\dd\theta\;\e^{-2\tdk}\,\e^{\ii\Psi(\tdk,\theta)},
\end{equation}
with
\begin{equation}\label{eq:Psi}
\Psi(\tdk,\theta)=w\tdk+\e^{-2\tdk}+2\sqrt\zeta\,\e^{-\tdk}\cos\theta .
\end{equation}
The integrals can be done in closed form and give Gordon's
Coulomb wave \cite{Gordon1928,Peters1974,DeguchiWatson1986},
\begin{equation}\label{eq:Fomega}
F_\omega(\bx')=N_w\,\hF\big(\ii\nu,1;\ii\omega\eta\big),\qquad N_w=\e^{\pi\nu/2}\,\Gamma(1-\ii\nu),
\end{equation}
which depends on the source position only through $\eta$. The meaning of the two remaining variables appears with the change
of variables
\begin{equation}\label{eq:xmap}
|\bx|=\sqrt{2/w}\;\e^{-\tdk},\qquad \arg\bx=\theta+\pi,\qquad y^2=\frac{2\zeta}{w}=\frac{\eta}{2M}.
\end{equation}
Then, $\e^{-2\tdk}\dd\tdk\,\dd\theta=\frac w2\dd^2x$ and $\Psi=w\Tf(\bx,\by)-\zeta-\nu\ln(w/2)$, so that
$F(w,y)\equiv\e^{\ii\nu\ln(w/2)}F_\omega$ is exactly the diffraction integral~\eqref{eq:kirchhoff}: the radius of the lens plane is
the DK time and its azimuth is the fibre angle (Appendix~\ref{App:B} gives both derivations in full). At fixed $\bx$ the frequency
enters only through $w=4M\omega$ multiplying the frequency-independent Fermat potential. Since $2\omega r/\nu=r/M$, the far-field Green function is
\begin{equation}\label{eq:Gfar_full}
G_\omega(r\bn,\bx')\to\frac{\e^{\ii\omega\Delta_0}}{4\pi r}\,F(w,y),\qquad \Delta_0=r-\bn\cdot\bx'+2M\ln\frac rM .
\end{equation}
No stationary-phase, eikonal, paraxial or thin-lens step has been taken: for a point mass and an observer at infinity the
thin-lens integral is exact, provided $y$ is defined by $y^2=\eta/2M$ rather than by projecting the source onto a plane.

\begin{figure}[t]\centering
\includegraphics[width=\textwidth]{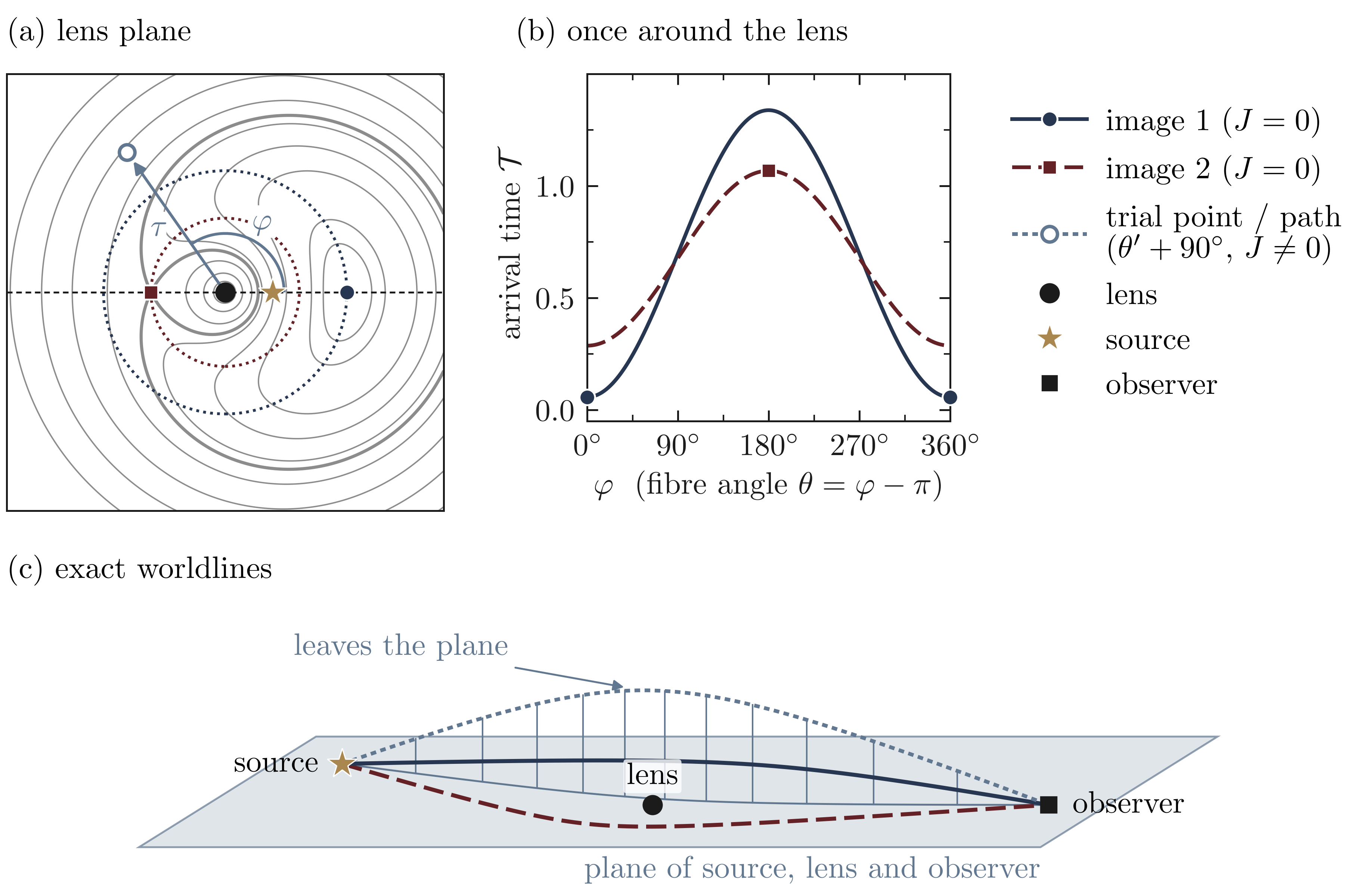}
\caption{The lens plane in worldline variables. (a) The distance from the lens is the DK time $\tdk$, and the clock angle
$\varphi=\theta+\pi$ is the KS fibre angle; grey lines are contours of $\Tf$. (b) $\Tf$ along the circles through the two images:
flat only at $0^\circ$ and $180^\circ$; image 2 is the highest point of its circle, hence a saddle. (c) Exact KS/DK worldlines: the
stationary points are planar Rutherford rays ($J=0$), while a trial path with $\theta$ moved by $90^\circ$ has $J\neq0$ and leaves
the plane.}\label{fig:azimuth}
\end{figure}

In these variables the stationary points of Sec.~\ref{sec:constraints} become explicit. With $u=\e^{-\tdk}$ they solve
\begin{equation}\label{eq:statpts}
\partial_\theta\Psi=-2\sqrt\zeta\,u\sin\theta=0,\qquad
\partial_\tdk\Psi=w-2u^2-2\sqrt\zeta\,u\cos\theta=0 .
\end{equation}
The fibre condition gives $\theta\in\{0,\pi\}$: both images lie on the line through the lens and the source, image 1 on
the source side ($\theta=\pi$) and image 2 on the opposite side ($\theta=0$). The DK-time condition is the radial lens
equation, which fixes their distances from the lens to the values $x_\pm$ of Eq.~\eqref{eq:images} (Fig.~\ref{fig:azimuth}).

\begin{figure}[h]\centering
\includegraphics[width=0.92\textwidth]{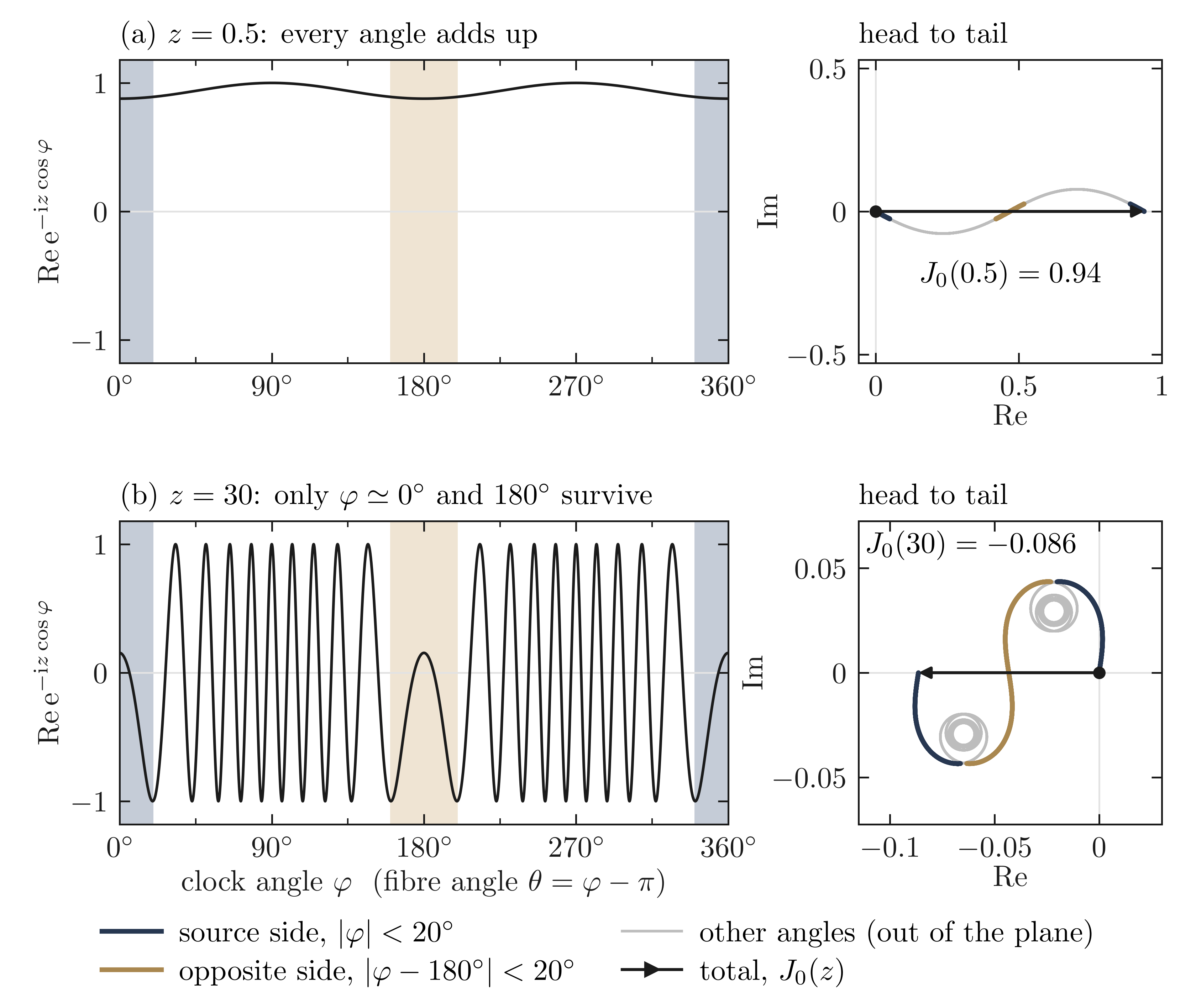}
\caption{Which clock angles contribute to $J_0(z)$, $z=wxy$. Left: real part of $\e^{-\ii z\cos\varphi}$ against the clock angle
$\varphi=\theta+\pi$. Right: the contributions added head to tail; blue and orange, within $20^\circ$ of the source side and of the
opposite side; grey, all other angles. For $z=0.5$ (top) nothing cancels; for $z=30$ (bottom) the grey stretches curl up and
cancel, and the straight runs at $0^\circ$ and $180^\circ$ are the two rays.}
\label{fig:phasors}
\end{figure}

This planarity, however, holds for the rays, not for the wave. At fixed DK time the integral over the fibre angle in
Eq.~\eqref{eq:far} can be done exactly,
\begin{equation}\label{eq:J0}
\frac1{2\pi}\int_0^{2\pi}\dd\theta\;\e^{\ii z\cos\theta}=J_0(z),\qquad z=2\sqrt\zeta\,\e^{-\tdk}=w\,x\,y=\frac{\omega\,\Delta t_\theta}{2},
\end{equation}
where $\Delta t_\theta=8Mxy$ is the difference in arrival time between the near and the far side of the lens along the
circle of radius $x$. All angles enter with the same weight, including the out-of-plane paths with $J\neq0$: the
constraint $J=0$ singles out the in-plane rays only once the integral is evaluated by stationary phase. If $z\lesssim1$,
the phase changes little around the circle, all angles add up and the wave reaches the observer from all around the
lens. If $z\gg1$, the contributions cancel except near $\theta=0$ and $\pi$, where the phase is stationary,
$J_0(z)\simeq(2/\pi z)^{1/2}\cos(z-\pi/4)$ with a relative error of order $1/(8z)$, and only the two rays survive
(Fig.~\ref{fig:phasors}).
\vspace{3pt}

Geometric optics therefore requires the phase to vary rapidly in both directions: around the circle, with curvature
$z_\pm=wx_\pm y$, and in the DK time, with curvature $w(1+x_\pm^2)$. Both grow with $w$, since $\Psi=w\Tf+{\rm const}$
with $\Tf$ independent of frequency, and since $x_\pm\simeq1$ near the axis the two conditions read
\begin{equation}\label{eq:GOcond}
w\gg1\qquad\text{and}\qquad wy\gg1 .
\end{equation}
Near the axis the error of geometric optics depends essentially on $wy$ alone, so a high frequency is not enough. When
both conditions hold, only the classical worldlines contribute, the planar Rutherford rays with $\Eosc=4M$ and $J=0$,
and the integral reduces to Eq.~\eqref{eq:FGO}, with Fermat phases, magnifications given by the Hessian and Morse
phases. This is the leading order of the expansion in $\hbar=1/\omega$, whose next terms are the beyond-GO
corrections~\cite{Takahashi2004,CarrilloGonzalez:2025gqm,Cusin:2019rmt}.

\section{The bound lens}\label{sec:bound}

As discussed in Section~\ref{sec:standard}, GW lensing is usually described by the diffraction integral~\eqref{eq:kirchhoff},
which rests on the assumptions collected in Table~\ref{tab:assumptions} \cite{Nakamura1998,NakamuraDeguchi1999,TakahashiNakamura2003}.

\begin{table}[h!]\centering\small
\begin{tabular}{>{\raggedright}p{2.6cm}>{\raggedright}p{5.0cm}>{\raggedright\arraybackslash}p{6.0cm}}
\toprule
Assumption & Physical content & Fails when\\
\midrule
eikonal, one-way & envelope slowly varying along the propagation; forward waves only & paths within a wavelength or a few $M$ of the lens; large-angle scattering\\
paraxial & small angles; source projected on a plane & source at a large angle from the lens axis\\
thin lens, far field & lens short compared with the distances & source or observer close to the lens\\
weak field & small deflection; no photon sphere & paths within a few $M$ of a compact object\\
static & $F(\omega)$ multiplies the spectrum & source or lens moves during the lensing delays\\
scalar & polarization parallel-transported & large scattering angles\\
\bottomrule
\end{tabular}
\caption{Assumptions behind the diffraction integral, their physical content and when they fail.}\label{tab:assumptions}
\end{table}

For a lens at cosmological distance from both source and observer all of them hold extremely well, which is why the
diffraction integral is the standard tool. However it is worth asking when they fail,
because this is where the worldline representation becomes necessary.
In order to do so, one first has to understand the physical and geometric meaning of each one of them. The least obvious is the eikonal step of
Eq.~\eqref{eq:ND_eikonal}, which reduces the second-order wave equation to a first-order, one-way evolution. To make it easier, it is worth analyzing it for a
point lens, in which case it can be tested exactly.

\begin{figure}[h!]\centering
\includegraphics[width=\textwidth]{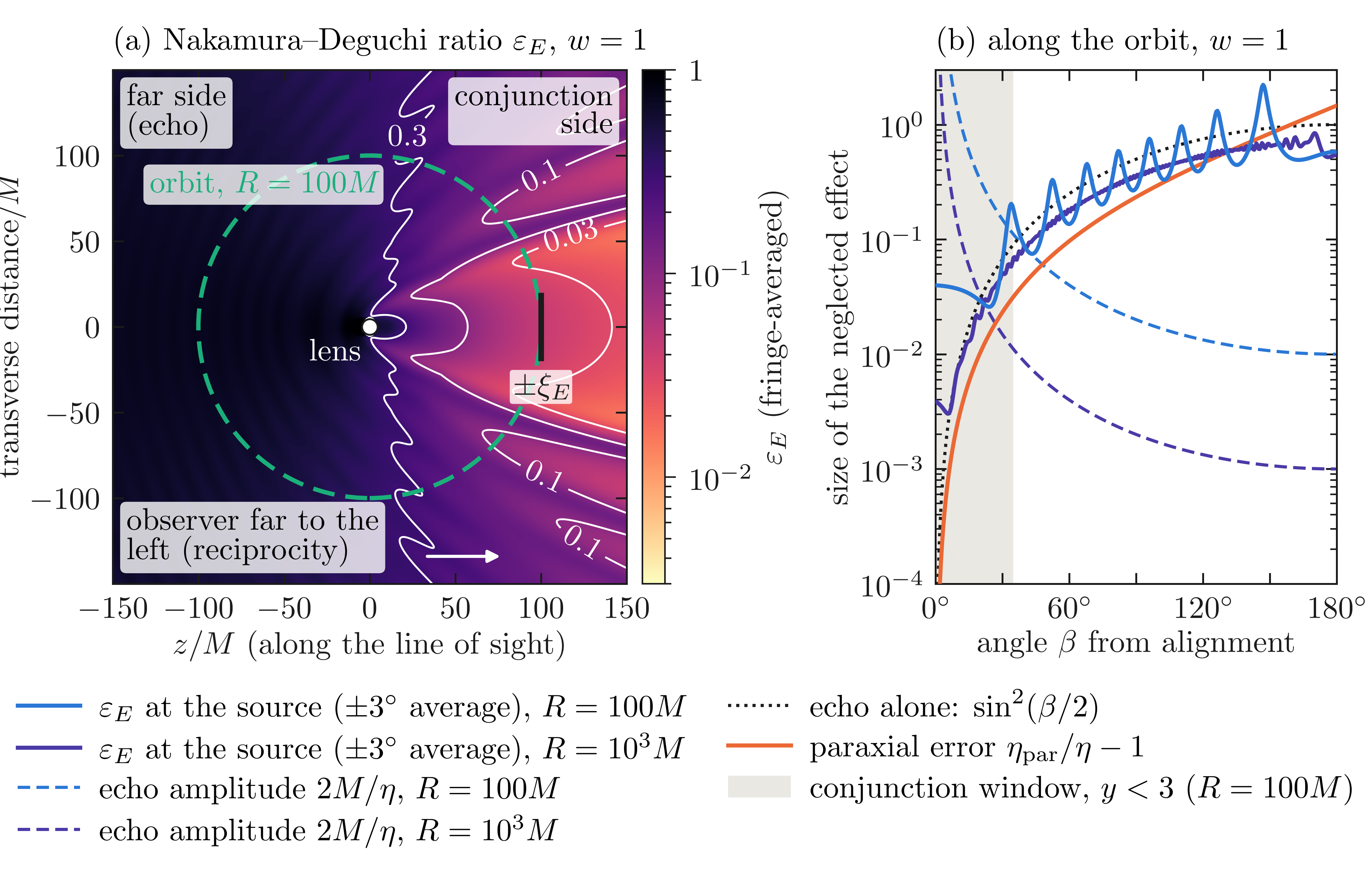}
\caption{Where the assumptions behind the diffraction integral fail around a point lens, $w=1$. (a) Exact ratio
$\varepsilon_E$ for the Coulomb wave, averaged over two fringes to remove interference spikes; by reciprocity the map also
describes a source at each point seen by a distant observer to the left. Dashed circle: an orbit of radius $R=100M$; white dot:
the lens; black bar: $\pm\xi_E$ at conjunction. The one-way reduction holds only in the forward cone behind the lens. (b) Along the
orbit, against the angle $\beta$ of the source from alignment: $\varepsilon_E$ at the source for $R=100M$ and $10^3M$, the same
quantity for the echo alone, the paraxial error and the echo amplitude. }
\label{fig:ndmap}
\end{figure}

\label{sec:eikonal_exact}For a plane wave incident on a point lens, the exact solution of Eq.~\eqref{eq:coulomb} is the
Coulomb wave, whose envelope $F$ is a confluent hypergeometric function of $\omega(r-z)$, with $z$ the propagation
coordinate. The eikonal step drops $\partial_z^2F$ against $2\ii\omega\,\partial_zF$, so it requires the ratio
\begin{equation}\label{eq:epsE}
\varepsilon_E\equiv\frac{|\partial_z^2F|}{|2\omega\,\partial_zF|}
\end{equation}
to be small. Evaluating it with the exact Coulomb wave, at distance $d$ from the lens and angle $\theta$ from the forward
direction, gives
\begin{equation}\label{eq:epsE_summary}
\varepsilon_E\simeq\frac{\bar\lambda}{d}+\frac{M}{d}+\mathcal O\Big(\sin^2\frac\theta2\Big),\qquad \bar\lambda=\frac1\omega .
\end{equation}
Each term is a separate condition. The first requires the wavelength to be short compared with the distance from the
lens, which is the scale over which the field of a point mass varies. The second, $M/d=|U|$, requires the field to be
weak. The third requires the wave to move forward, because a wave scattered backwards along $z$ cannot be described by
an equation that evolves only forward in $z$. These conditions must hold where the relevant paths pass the lens, at
distances of the order of the Einstein radius $\xi_E$, or of the Fresnel radius when $w\lesssim1$. There the field is
weak and the deflection small, so only the first term matters, and $\bar\lambda\ll\xi_E$ turns the eikonal step into a
lower bound on the frequency,
\begin{equation}\label{eq:fE}
f\gg f_E\equiv\frac{1}{2\pi\xi_E}\,.
\end{equation}
Figure~\ref{fig:ndmap} shows $\varepsilon_E$ around a point lens at $w=1$. The one-way reduction is accurate only in a cone
behind the lens, around the forward direction. Towards the lens the first two terms of Eq.~\eqref{eq:epsE_summary} grow: at
$w=1$ the reduced wavelength is $\bar\lambda=4M$, so that on the axis $\varepsilon_E\simeq0.04$ at $100M$ behind the lens and
it increases as the lens is approached. Away from the axis the angular term takes over, and on the far side $\varepsilon_E$ is of
order unity, because there the field contains the wave scattered backwards by the lens, which an equation that evolves only
forward in $z$ cannot describe. For a source on an orbit of radius $R$ the ratio at the source therefore grows from
$\simeq\bar\lambda/R$ at conjunction to order unity at opposition, following the contribution $\sin^2(\beta/2)$ of the echo,
while the paraxial excess path is wrong by up to $\pi^2/4-1\simeq1.5$ [Fig.~\ref{fig:ndmap}(b)]. The ratio $\varepsilon_E$
measures the size of the term dropped from the equation, evaluated on the exact solution, and not the error of the resulting
amplification factor. For a point mass the latter survives even where $\varepsilon_E\sim1$. The exact wave depends on
position only through $r-z$ and is therefore constant on paraboloids with focus at the lens, each of which reaches the forward
cone where the one-way reduction holds. The result of the diffraction integral obtained there then extends to every point,
provided the offset is defined by the excess path, $y^2=\eta/2M$ (Secs.~\ref{sec:lensplane} and~\ref{sec:bound_far}). This is
a property of the $1/r$ potential alone: for any other lens the dark regions of the map would turn into actual errors.

\label{sec:NR}The meaning of the eikonal step is clearest in field-theory language. If the distance along the
propagation plays the role of time, the frequency that of a mass and the transverse wavenumber that of a momentum, the
Helmholtz equation becomes a Klein--Gordon equation, and the eikonal step is its non-relativistic
limit~\cite{Schulman1984,RevModPhys.38.483,Palmer1979}. Just as the non-relativistic limit discards the antiparticles,
the eikonal step keeps only the wave moving forward and discards the one moving backward.  A backward wave can arise
in two ways. The first is a classical turning point, where a ray itself reverses its direction; this requires a strong
deflection, i.e.\ a path within a few gravitational radii of a very compact object, outside the linearized field. The
second is backscattering by diffraction, which needs no ray to turn back: the $1/r$ potential has Fourier components at
every momentum transfer, including the one that reverses the wave, so it scatters part of the wave in every direction,
backwards included. This backscattered wave is the echo of a source close to its lens, the scattered part of the exact
Coulomb wave (Sec.~\ref{sec:bound_far}). Its relative amplitude decreases with the distance of the source from the
lens, so it is negligible on a cosmological line of sight but not for a source close to its lens.

\label{sec:timedep}A nearby source also violates the static assumption. The amplification factor multiplies the
spectrum only if the configuration is invariant under time translations, so that each worldline contributes a delayed
copy of the emitted signal. If the source moves during the spread of delays of the contributing worldlines, each of them
samples it at a different emission time, position and Doppler factor, and no amplification factor exists. The usual
treatment, which evaluates the static amplification factor at the instantaneous position of the
source~\cite{DOrazioLoeb2020,Yu2021}, must then be replaced by the exact time-domain Green function, in which each
worldline has its own retarded time. The remaining assumptions are milder: the paraxial projection of the source onto a
plane is accurate near the lens axis but wrong by a factor of order one at large angles, the
thin-lens corrections are suppressed by the ratio of the Schwarzschild radius to the distances, and the weak-field and
scalar approximations fail only for paths within a few gravitational radii of the lens or at large scattering angles.

\label{sec:synthesis}All these effects are negligible on a cosmological line of sight, but they appear together when the
source orbits its lens (Fig.~\ref{fig:ndmap}): near conjunction the paths pass close to the lens, on the rest of the
orbit the source is seen at large angles and its backscattered echo interferes with the direct wave, and along the whole
orbit the source moves during the lensing delays. This \emph{bound lens} is therefore the natural setting for the exact
worldline treatment.

\subsection{The bound lens: triple systems }\label{sec:bound_why}
Normally a gravitational lens is a mass that happens to sit between a distant source and us, with all three objects billions of light-years apart. In a bound lens, or triple system, the source orbits the lens \cite{Kocsis2013,DOrazioLoeb2020,Yu2021,Pijnenburg2024,Ubach:2025dob}. A typical case is a binary of small black holes circling a much heavier black hole, for example in a galactic nucleus. Once per orbit the source passes behind the lens as we see it, and its signal is lensed. For the rest of the orbit it is beside or in front of the lens \cite{Pijnenburg2024, Cardoso:2026ugm,Yu2021,Grespan:2023cpa}.

\label{sec:bound_model}With this idea in mind, we consider a gravitational-wave source on a circular orbit around a point lens. The lens has mass $M$ and the orbit has radius $R$, so that the
source moves with angular frequency $\Omega=(M/R^3)^{1/2}$ and speed $v=(M/R)^{1/2}$. A distant observer sees the system along the
unit vector $\bn$ that points from the lens to the observer, and the orbital plane is inclined by an angle $i$ with respect to the
plane of the sky. Once per orbit the source passes behind the lens, at superior conjunction, and we measure the orbital phase
$\Phi=\Omega t$ from there; half an orbit later, at opposition, the source passes in front of the lens. For an edge-on orbit,
$i=90^\circ$, the source goes exactly behind the lens (Fig.~\ref{fig:bound_coords}).

This configuration combines all the effects that the diffraction integral leaves out. The source is close to the lens, so the
waves reach the observer after being deflected by every angle and part of them is scattered back; and the source moves while the
signal crosses the lens. Our goal in this section is to compute the received signal exactly within a well-defined model, and then
to use the exact result to identify, worldline by worldline, what each standard approximation keeps and where along the orbit it
fails.

For the sake of simplicity, we consider a scalar wave propagating in the first order metric of Eq.~\eqref{eq:metric} with a point mass lens potential, meaning
\begin{equation}
  g_{00}=-(1-2M/r) \,,\qquad  g_{ij}=(1+2M/r)\delta_{ij}\,.
\end{equation}
 Moreover, we assume the lens to be static, which requires the mass $m$ of the source to be negligible, $m\ll M$, so
that the reflex motion of the lens, of speed $(m/M)v$, can be ignored. A static field makes the problem invariant under time
translations: the frequency is conserved along every route from the source to the observer, and the Green function depends on
$t-t'$ only. We emphatize that this is the only static assumption, since the source itself may move arbitrarily. Finally, we require  that the source radiates
isotropically. The linearized point lens is an idealization: at distances of $10^2$--$10^3M$ from a black hole the post-Minkowskian
corrections to the deflection~\cite{CarrilloGonzalez:2025gqm} are not negligible, and waves scattered through large angles
probe the strong field near the photon sphere, which requires black-hole perturbation
theory~\cite{Pijnenburg2024,Chan:2025wgz,Saketh:2025cwf}. We estimate these corrections, together with those due to the spin
of the wave and to the emission pattern of the source, in Sec.~\ref{sec:bound_beyond}. 

The choice of a point lens deserves a comment. For a static point lens the worldline and the diffraction-integral
descriptions agree: the exact far-field Green function of Sec.~\ref{sec:lensplane} has the form of the point-lens
amplification factor~\eqref{eq:Fwy} at every position of the source, provided the offset is defined by the excess path,
$y^2=\eta/2M$. This agreement, however, is entirely due to the form of the Coulomb potential and does not mean that the
approximations behind the diffraction integral hold. Because the $1/r$ potential separates in parabolic coordinates, the
exact wave depends on position only through $r-z$ and is therefore constant on paraboloids with focus at the lens. Each of
them reaches the forward region behind the lens, where the eikonal, paraxial and thin-lens steps are accurate, and the value
obtained there extends to every point of the paraboloid, including those near the lens and at large angles, where the same
steps fail (Fig.~\ref{fig:ndmap}). For any other lens, and already for the second-order Schwarzschild field, no such property
exists. We use the point lens because it is the case that the Duru--Kleinert gauge solves exactly (Sec.~\ref{sec:DK}), so
that the treatment of the bound lens rests on an exact propagator rather than on an approximation. Even for this lens,
however, the bound configuration contains effects that the diffraction integral does not consider at all. The
amplification factor describes a static configuration, whereas the source moves during the lensing delays. Each worldline
then delivers the signal emitted at its own retarded time and with its own Doppler factor: the echo received at a given time
left the source earlier than the direct wave received with it and, on a circular orbit, carries no orbital Doppler shift,
so that no single amplification factor describes the received wave. The offset must be defined by the excess path along the
whole orbit, since the small-angle offset misplaces the echo on the far side and a projected one would make a source in
front of the lens appear aligned with it. Finally, the echo itself is a wave scattered backwards by the lens, which the
one-way equation excludes by construction and which the closed form contains only through the parabolic symmetry. These are
the effects captured by the exact time-domain signal of Sec.~\ref{sec:bound_time} and missed by the approximations of
Sec.~\ref{sec:bound_models}.

\subsection{A static source: direct wave and echo}\label{sec:bound_far}

\begin{figure}[h!]\centering
\includegraphics[width=\textwidth]{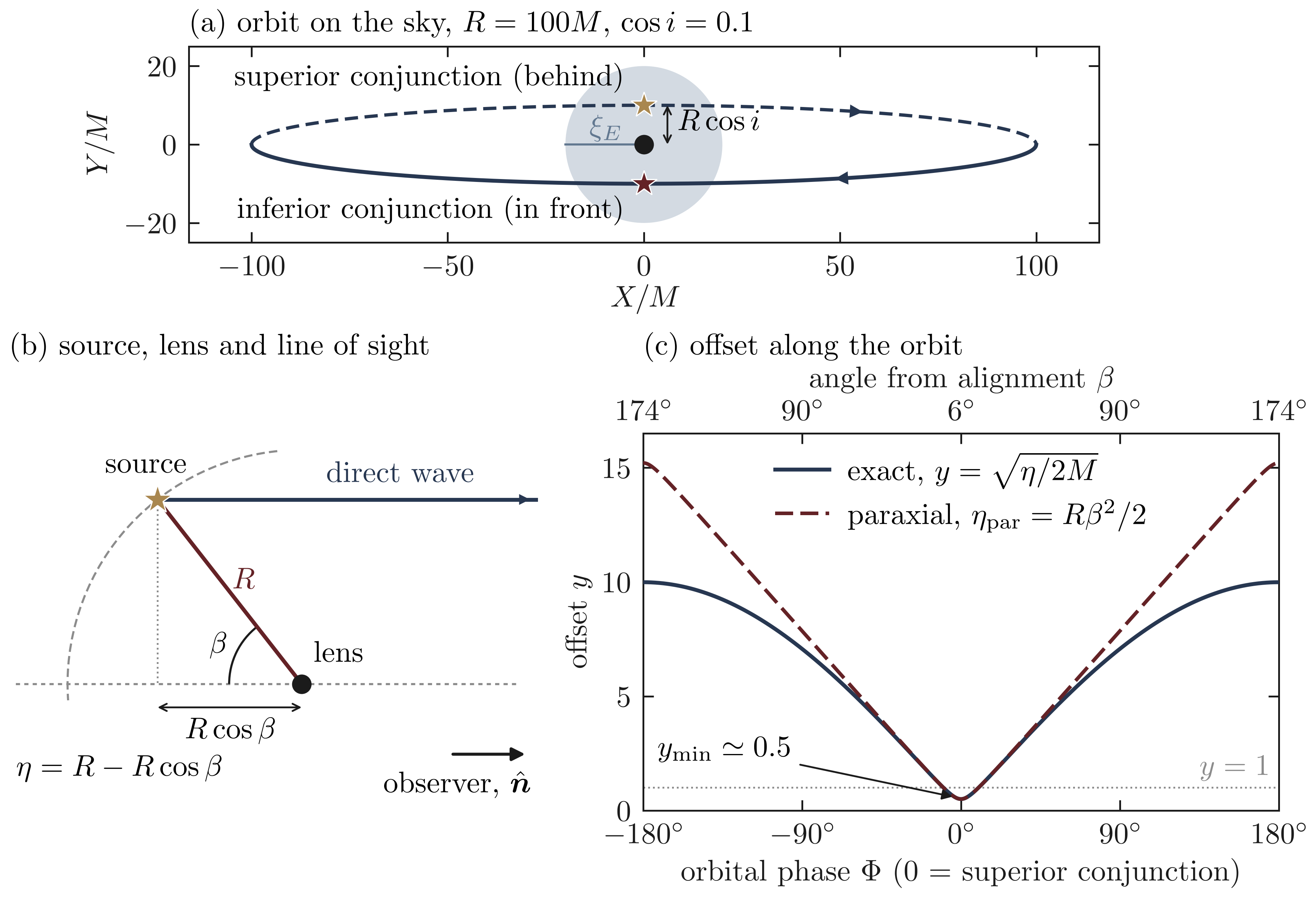}
\caption{Geometry of the bound lens. (a) The orbit on the sky for $R=100M$ and $\cos i=0.1$: an ellipse of semi-axes $R$ and
$R\cos i$; at superior conjunction the source passes $R\cos i=10M$ from the lens, i.e.\ $y_{\min}\simeq0.5$ Einstein radii. (b) The
plane of source, lens and line of sight: the angle $\beta$, the R\o mer advance $R\cos\beta$ of the direct wave and the excess path
$\eta=R-R\cos\beta$ of the route via the lens. (c) The exact offset $y=\sqrt{\eta/2M}$ and the paraxial offset along the orbit,
together with $\beta$. }\label{fig:bound_coords}
\end{figure}

We begin with a source frozen at a point $\bx'$ of its orbit, the situation described by the amplification factor. Two quantities
fix its position with respect to the line of sight. The first is the projection $\bn\cdot\bx'=-R\sin i\cos\Phi$: a source farther
than the lens along the line of sight has $\bn\cdot\bx'<0$, and its direct wave arrives later by the R\o mer delay $\Delta_R=-\bn\cdot\bx'$; when the source is in front of the lens, it arrives earlier by the same amount. For gravitational-wave sources in hierarchical triples this delay is a known modulation of the signal~\cite{Meiron2017}. The second is the angle $\beta$ between $\bx'$ and $-\bn$, given by $\cos\beta=\sin i\cos\Phi$, which measures
how far the source is from perfect alignment behind the lens. Thus $\beta=0$ is exact alignment behind the lens, $\beta=\pi$ exact
alignment in front of it, and a nearly edge-on orbit comes within $\beta_{\min}\simeq\cos i$ of alignment once per orbit. The route
from the source to the observer via the lens is longer than the direct route by
\begin{equation}\label{eq:eta_bound}
\eta=r'+\bn\cdot\bx'=R(1-\cos\beta),
\end{equation}
because the direct wave needs a distance $R\cos\beta$ to reach the wavefront through the lens, while the route via the lens needs
$R$ [Fig.~\ref{fig:bound_coords}(b)]. This excess path ranges from $R(1-\sin i)$ at conjunction to $R(1+\sin i)$ at opposition.

For an observer at infinity, the exact far-field Green function~\eqref{eq:Gfar_full} separates into four factors,
\begin{equation}\label{eq:far_bound}
G_\omega(r\bn,\bx')\to\frac{\e^{\ii\omega r}}{4\pi r}\e^{2\ii M\omega\ln(r/M)}\e^{-\ii\omega\bn\cdot\bx'}F(w,y)\, .
\end{equation}
The first factor is the free propagation to the observer, the second the Shapiro delay accumulated on the long leg from the lens
to the observer, and the third the R\o mer phase of the source position. The first three are pure delays, linear in $\omega$, and
everything specific to lensing is in the amplification factor $F(w,y)$. This is the diffraction integral~\eqref{eq:kirchhoff},
whose closed form is Eq.~\eqref{eq:Fwy}, evaluated at the offset
\begin{equation}\label{eq:y_bound}
y=\Big(\frac{\eta}{2M}\Big)^{1/2}.
\end{equation}
The amplification factor therefore depends on the position of the source only through
$\eta=r'+\bn\cdot\bx'$, the extra length of the route via the lens with respect to the direct one. This
follows from reciprocity. Because the lens is static, the wave operator Eq~\eqref{eq:helmholtz} is
symmetric, and the propagator is unchanged when source and observer are exchanged,
$G_\omega(\bx,\bx')=G_\omega(\bx',\bx)$. A source placed at the observer, far along $\bn$, sends towards the
lens a plane wave traveling along $-\bn$, so $F$ is also the amplitude, at the position of the source,
of a plane wave coming from the direction of the observer. In the field of a point mass this is the
Coulomb wave, $\e^{\ii\omega z}\,\hF\big(\ii\nu,1;\ii\omega(r-z)\big)$ with $z=-\bn\cdot\bx$ measured along the
direction of propagation~\cite{Gordon1928}. Its envelope depends on position only through the parabolic
coordinate $r-z$, which at the source equals $\eta$, so $F$ is constant on paraboloids of revolution with
focus at the lens and axis along the line of sight. The offset $y$ is thus defined at every orbital phase, without projecting the source onto a plane. Near conjunction, where
$\eta\simeq R\beta^2/2$, it reduces to the usual offset $y\simeq\beta R/\xi_E$ in units of the Einstein radius $\xi_E=(4MR)^{1/2}$ of a
lens at distance $R$ from the source. Its minimum, reached at conjunction, is $y_{\min}\simeq(R/4M)^{1/2}|\cos i|$, e.g.\ $0.5$ for
$R=100M$ and $\cos i=0.1$. In what follows we call \emph{conjunction window} the part of the orbit with $y<3$, which covers about a
fifth of the orbit for $R=100M$ and 5--6\% of it for $R=10^3M$, and \emph{far side} the part with $\beta>\pi/2$.

Away from conjunction the excess path is many wavelengths, $\zeta\equiv\omega\eta\gg1+\nu^2$ with $\nu=w/2$, and the asymptotic expansion
of $F$ separates two waves,
\begin{equation}\label{eq:Fasym}
F\simeq\underbrace{y^{-\ii w}\Big[1-\frac{\ii\nu^2}{\zeta}+\ldots\Big]}_{\text{direct wave}}
+\underbrace{\nu^{\ii\nu}N_w\,\frac{\e^{\ii\zeta}(\ii\zeta)^{\ii\nu-1}}{\Gamma(\ii\nu)}\big[1+\mathcal O(\zeta^{-1})\big]}_{F_{\rm sc}:\ \text{echo}} ,
\end{equation}
with $N_w$ defined in Eq.~\eqref{eq:Fomega}. The direct wave carries the Shapiro phase $-w\ln y$ at every angle, together with the
bending correction $-\nu^2/\zeta$. The scattered wave, or \emph{echo}, has relative amplitude $|F_{\rm sc}|=\nu/\zeta=2M/\eta$ and the extra
phase $\e^{\ii\zeta}$: it is the wave that travelled from the source to the lens and was scattered towards the observer, arriving
later by the excess path $\eta$ \cite{Kocsis2013}. Near alignment it becomes the second image. Along the orbit its amplitude ranges
from $2M/R$ at quadrature to $M/R$ at opposition, i.e.\ 1--2\% for $R=100M$.

The static amplification factor along the orbit follows directly. For $w=1$, $R=100M$ and $\cos i=0.1$, $|F|$ peaks at $1.70$ at
superior conjunction, where $y=y_{\min}=0.50$ [Fig.~\ref{fig:bound}(a)]. The peak is limited by diffraction: the on-axis value is
$|F(1,0)|=1.81$, whereas geometric optics gives $\sqrt{\mu_+}+\sqrt{\mu_-}=2.03$ at $y=0.5$ and diverges for an aligned orbit. Away from
conjunction $|F|$ oscillates within $1\pm2M/\eta$ as the echo interferes with the direct wave, while its phase follows the Shapiro
phase.

The worldline picture of Sec.~\ref{sec:lensplane} gives a meaning to each part of this result. Near conjunction the lens-plane
variable $\bx$ coincides with the physical point, in units of $\xi_E$, where a path crosses the plane through the lens
perpendicular to $\bn$. On the far side there is no such plane between source and observer, but the variables keep their worldline
meaning: $|\bx|$ measures the DK time of a worldline and $\arg\bx$ its fibre angle. The second image, at $x_-\simeq1/y\ll1$, is then
the echo, a worldline with a very long DK time that passes very close to the lens. In the linearized field it is a Rutherford
orbit deflected by $\beta$, with impact parameter $b=2M\cot(\beta/2)$, which is smaller than $2M$ for $\beta>\pi/2$. Such a deflection
is a strong-field effect, and the far-side echo is therefore the least reliable part of the linearized model
(Sec.~\ref{sec:bound_beyond}).

In this language, each approximation used in the literature is a stationarity assumption with a definite physical content, and
each fails in a definite part of the orbit. Stationarity in the DK time puts every worldline on the energy shell of the lens and
fixes the distance of the images from it (the radial lens equation); stationarity in the fibre angle makes every worldline planar
and puts the images in the plane of source, lens and observer. Imposing both gives geometric optics. Since $y\simeq\beta(R/4M)^{1/2}$ near conjunction, GO fails
within
\begin{equation}\label{eq:betaGO}
\beta_{\rm GO}\simeq\frac{2}{w}\Big(\frac MR\Big)^{1/2}
\end{equation}
of conjunction, e.g.\ $0.2$\,rad for $w=1$ and $R=100M$.

A simpler approximation retains a single straight worldline and assigns to it the DK phase, corresponding to its Shapiro delay. This is the eikonal
approximation in the sense of high-energy scattering \cite{CarrilloGonzalez:2025gqm}, and it gives $F_{\rm SL}=y^{-\ii w}$, the
first term of the direct wave~\eqref{eq:Fasym}. Since $|F_{\rm SL}|=1$, there is no focusing, and the model misses the bending
correction, the second image and the echo. It requires both corrections to be small, $\zeta\gg\nu(1+\nu)$, i.e.\ $\eta\gg2M(1+\nu)$, and
therefore fails within
\begin{equation}\label{eq:betaeik}
\beta_{\rm eik}\simeq\Big[\frac{4M(1+\nu)}{R}\Big]^{1/2}
\end{equation}
of conjunction ($0.24$\,rad for $w=1$, $R=100M$), and misses the echo everywhere. Finally, the paraxial geometry replaces $\eta$ by
its small-angle form $\eta_{\rm par}=R\beta^2/2$, which is exact near conjunction but wrong by up to a factor $\pi^2/4$ on the far side.
None of these approximations concerns the motion of the source, which we now include.

\subsection{A moving source: the exact signal}\label{sec:bound_time}\label{sec:bound_signal}

When the source moves, the amplification factor is no longer defined. The lens, however, is static, so the retarded Green
function depends only on $t-t'$, and for a source of strength $q(t)$ on the trajectory $\bx_s(t)$ the exact field is
\begin{equation}\label{eq:fieldmoving}
\varphi(\bx,t)=\int\dd t'\;q(t')\,G_R\big(\bx,t;\bx_s(t'),t'\big)\,.
\end{equation}
 For an observer at infinity, $G_R$ is
the time-domain Green function
\begin{equation}\label{eq:GRrho}
G_R(r\bn,t;\bx',t')=\frac{1}{8\pi^2r}\,\frac{\partial}{\partial t}\,\rh\Big(\frac{t-t'-\Delta_0}{4M};y\Big),\qquad
\Delta_0=r-\bn\cdot\bx'+2M\ln\frac rM\,,
\end{equation}
in which the lens enters only through the delay density~\eqref{eq:rho_std}.We remove the delay common to all worldlines by defining the arrival time
$T\equiv t-r-2M\ln(r/M)$, and we normalize the signal as $h(T)\equiv4\pi r\,\varphi(r\bn,t)$. Inserting
Eq.~\eqref{eq:GRrho} into Eq.~\eqref{eq:fieldmoving} and calling now $t'=t$ we get
\begin{equation}\label{eq:hT}
h(T)=\frac{1}{2\pi}\,\frac{\dd}{\dd T}\int\dd t\;q(t)\;
\rh\Big(\frac{T-t-\Delta_R(t)}{4M};\,y(t)\Big),
\end{equation}
where $\Delta_R(t)=-\bn\cdot\bx_s(t)$ is the R\o mer delay and $y(t)$ of Eq.~\eqref{eq:y_bound} are now evaluated along the orbit.
This formula holds for any time dependence of the source, for any orbit, and with
no assumption about time scales. Physically, Eq.~\eqref{eq:hT} treats each instant of
emission as a flash sent from the position the source had at that instant. The lens turns each flash into
its response $\rh/2\pi$, shifted by the R\o mer delay $\Delta_R(t)$ and evaluated with the offset $y(t)$ of
that moment, and the signal received at $T$ is the sum of these responses over the whole source history.
Equation~\eqref{eq:hT} is thus a convolution whose kernel changes with the emission time, which is why no
single amplification factor describes a moving source.

\begin{figure}[h!]\centering
\includegraphics[width=\textwidth]{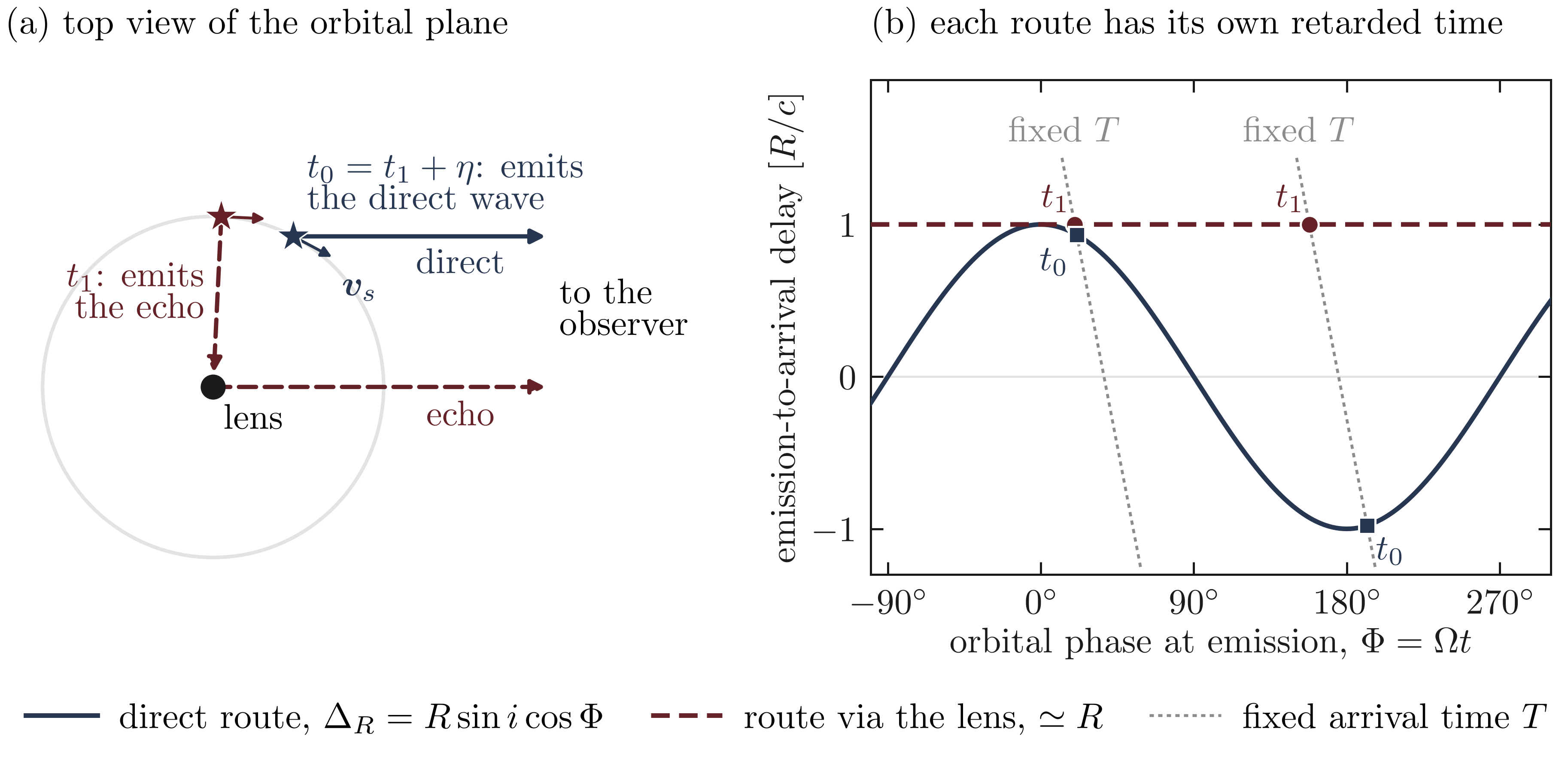}
\caption{Each route has its own retarded time (schematic: edge-on orbit, speed exaggerated to $v=0.3$). (a) Top view of the orbital
plane, observer to the right. The signal arriving at $T$ was emitted as the echo at $t_1$ and as the direct wave at $t_0=t_1+\eta$; in
between the source has moved. The echo leaves towards the lens, perpendicular to the orbital velocity, and is not Doppler shifted.
(b) Delay between emission and arrival against the emission phase: R\o mer delay $\Delta_R=R\sin i\cos\Phi$ for the direct route, $R$ for
the route via the lens. A line of slope $-1$ (fixed arrival time) cuts each route at its own emission time, and the slope of each
route gives its Doppler factor.}\label{fig:retarded}
\end{figure}

To make this more explicit, we can look at two limits of the system. With no lens, $\rh/2\pi$ is a unit step and Eq.~\eqref{eq:hT} reduces to the unlensed signal
\begin{equation}\label{eq:h0}
h_0(T)=q(t_0)\,\frac{\dd t_0}{\dd T},\qquad T=t_0+\Delta_R(t_0),
\end{equation}
which contains the R\o mer delay of the emission time $t_0$ and the Doppler factor $\dd t_0/\dd T=(1-\bn\cdot\bm v_s)^{-1}$, with
$\bm v_s$ the velocity of the source. To see how the lens acts, we write $\rh$ as the lens-plane integral
$\rh(\tauh;y)=\int\dd^2x\,\delta\big(\tauh-\Tf(\bx,\by)\big)$. For each lens-plane point $\bx$, i.e.\ for each worldline, a signal emitted at
time $t$ arrives at
\begin{equation}\label{eq:Tx}
T_{\bx}(t)=t+\Delta_R(t)+4M\,\Tf\big(\bx,\by(t)\big),
\end{equation}
and the inverse function $t_{\bx}(T)$ is the retarded time of that worldline. The $t$-integral in Eq.~\eqref{eq:hT} then collapses
onto it,
\begin{equation}\label{eq:hTx}
h(T)=\frac{2M}{\pi}\,\frac{\dd}{\dd T}\int\dd^2x\;q\big(t_{\bx}(T)\big)\,\frac{\dd t_{\bx}}{\dd T}.
\end{equation}
Each worldline therefore delivers the signal emitted at its own retarded time, with its own Doppler factor $\dd t_{\bx}/\dd T$. Its
delay $4M\Tf$ contains the geometric path and, through the term $-4M\ln|\bx|$, the Shapiro delay, i.e.\ $4M$ times its DK time. For a
static source with $q=\e^{-\ii\bar\omega t}$, all worldlines share the same retarded geometry and Eq.~\eqref{eq:hTx} gives back
$h=q(T-\Delta_R)\,F(w,y)$. Figure~\ref{fig:retarded} illustrates the difference between the direct wave and the echo.

\section{Comparison with the existing results}
\label{sec:bound_models}

The exact formula~\eqref{eq:hTx} makes it possible to derive each approximate model and to see what it neglects. Most treatments of
an orbiting source evaluate the static amplification factor at the instantaneous position of the source \cite{DOrazioLoeb2020,Yu2021}.
This model follows from Eq.~\eqref{eq:hTx} for a monochromatic source, $q=\e^{-\ii\bar\omega t}$, with two assumptions:
the lens geometry is frozen at the direct retarded time, $\by(t)\to\by(t_0)$, and the R\o mer delay is linearized around $t_0$. The
retarded time of each worldline then becomes $t_{\bx}\simeq t_0-4M\Tf\,\dd t_0/\dd T$, and the lensing modulation is
\begin{equation}\label{eq:adiab}
\Lc_{\rm ad}(T)\equiv\frac{h(T)}{h_0(T)}\simeq F\big(w_{\rm loc},y(t_0)\big),\qquad w_{\rm loc}=\frac{4M\bar\omega}{1-\bn\cdot\bm v_s(t_0)},
\end{equation}
the static amplification factor at the retarded position of the source and at the Doppler-shifted frequency received. It amounts
to assuming that the source configuration is invariant under time translations during the lensing delays. We know though that this assumption does not alway hold. 
Indeed, this approach misses a transit effect near conjunction. The worldlines in the diffraction spot have delays that differ by
$4M\Delta\tauh\simeq8My$, and near conjunction the offset changes at the rate $\dd y/\dd t\simeq\Omega(R/4M)^{1/2}=1/(2R)$. During the
spread of delays $y$ therefore changes by $\delta y\simeq4My/R$, and the amplification factor by $\delta\ln F\sim w\,\delta y$, so that for
$w\gtrsim1$ the adiabatic error is $\approx(1$--$2)\,wM/R$.
Moreover, on the far side it misses the fact that the echo is not Doppler
shifted. On a circular orbit
\begin{equation}\label{eq:echodelay}
\Delta_R+\eta=R\sin i\cos\Phi+R\,(1-\sin i\cos\Phi)=R
\end{equation}
exactly, so the echo arrives at a fixed delay $R$ after emission, up to a slowly varying logarithm. It leaves the source towards the
lens, a direction in which a circular orbit has no velocity, whereas the direct wave carries the R\o mer delay and the Doppler
factor (Fig.~\ref{fig:retarded}). The adiabatic model imposes the Doppler shift of the direct wave on the echo, and its echo phase is
in error by $(\omega_{\rm loc}-\bar\omega)\eta$. Along the orbit this reaches $\simeq0.33\,w(R/M)^{1/2}$\,rad at $\Phi\simeq2\pi/3$, i.e.\
$3.3w$\,rad for $R=100M$ and $10w$\,rad for $R=10^3M$. For $w(R/M)^{1/2}\gtrsim3$ the adiabatic echo is uncorrelated from the true one,
and the error is of the order of the echo amplitude itself.
\subsection{A periodic source and the lensing modulation}\label{sec:comb}

\begin{figure}[h!]\centering
\includegraphics[width=\textwidth]{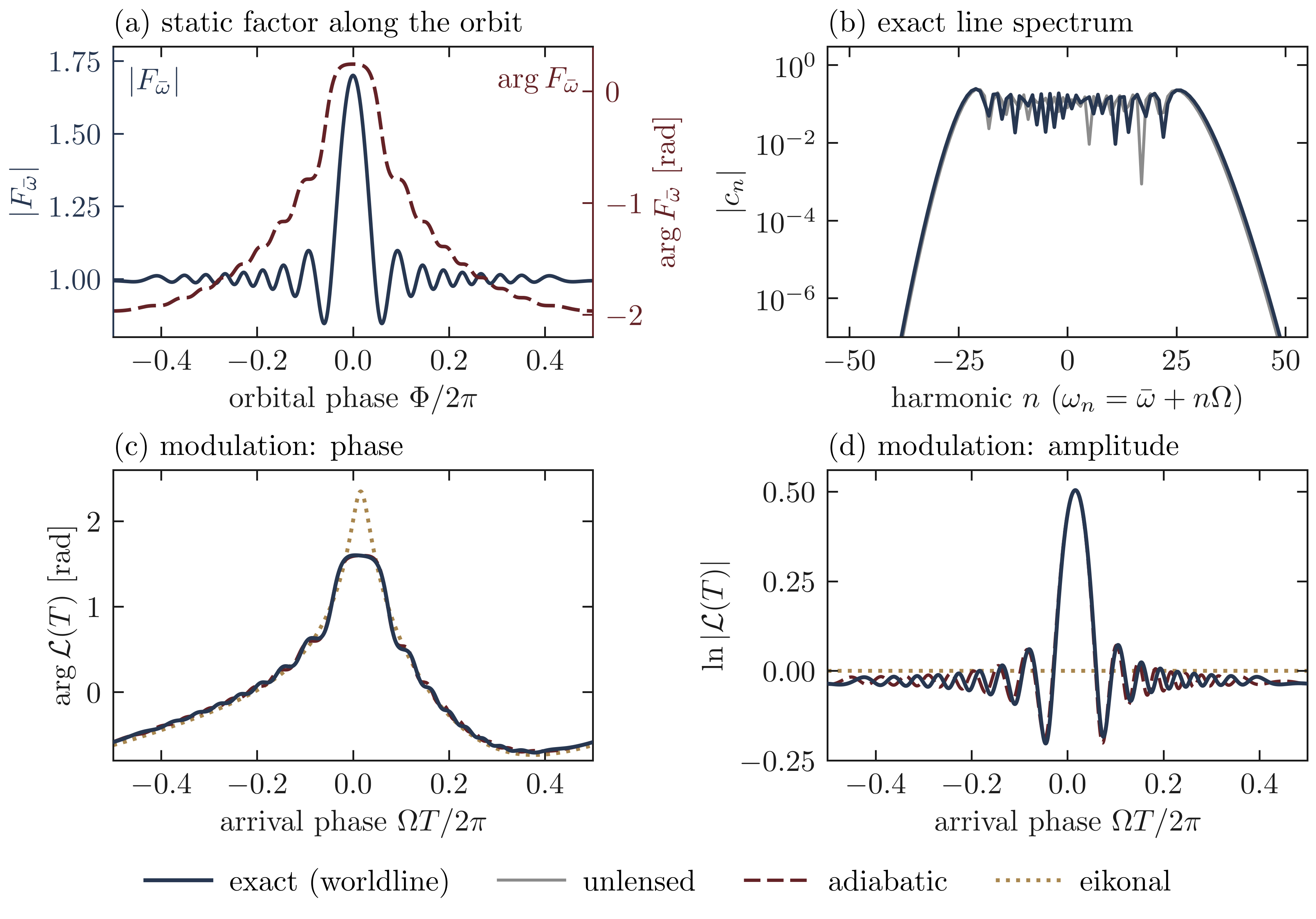}
\caption{Exact bound-lens result for $w=1$, $R=100M$ and $\cos i=0.1$. (a) Static amplification factor $F(w,y)$ along the orbit,
modulus (top) and phase (bottom); the inset shows the echo interference on the far side within the envelope $1\pm2M/\eta$. (b) Line
spectrum of the exact comb~\eqref{eq:comb}, lensed and unlensed. (c) Phase and (d) log-amplitude of the lensing
modulation~\eqref{eq:Lmod}, with the orbit average removed, against the arrival phase; the insets zoom on conjunction. Solid: exact;
dashed: adiabatic model~\eqref{eq:adiab}; dotted: eikonal model. }\label{fig:bound}
\end{figure}

The time-domain formula~\eqref{eq:hT} holds for any source. For the simplest one, a monochromatic source $q=q_0\e^{-\ii\bar\omega t}$ on
a circular orbit, the same signal can also be computed in the frequency domain, which gives both a transparent spectral picture and
an independent check. We insert the far-field propagator~\eqref{eq:far_bound} into Eq.~\eqref{eq:fieldmoving} and use the arrival
time $T$, which removes the delays common to all emission times. The signal becomes
\begin{equation}\label{eq:hfreq}
h(T)=q_0\int\frac{\dd\omega}{2\pi}\,\e^{-\ii\omega T}\int\dd t\;\e^{\ii(\omega-\bar\omega)t}\,g_\omega(t)\,,\qquad
g_\omega(t)=\e^{-\ii\omega\bn\cdot\bx_s(t)}\,F\big(4M\omega,y(t)\big)\,,
\end{equation}
where $g_\omega(t)$ collects the R\o mer phase and the amplification factor at the position of the source at the emission time $t$.
On a circular orbit $g_\omega$ is periodic with the orbital period $P=2\pi/\Omega$. Expanding it in a Fourier series, the $t$-integral
becomes a sum of delta functions at $\omega=\bar\omega+n\Omega$, and the signal collapses onto a comb of lines,
\begin{equation}\label{eq:comb}
h(T)=q_0\sum_nc_n\,\e^{-\ii\omega_nT},\qquad
c_n=\frac1P\int_0^P\dd t\;\e^{\ii n\Omega t}\,\e^{-\ii\omega_n\bn\cdot\bx_s(t)}\,F\big(4M\omega_n,y(t)\big)\,, 
\end{equation}
with $\omega_n=\bar\omega+n\Omega$.
Without the lens, $F=1$ and the coefficients are Bessel functions, $c_n^{(0)}=\ii^nJ_n(\omega_nR\sin i)$: the R\o mer phase spreads the
line $\bar\omega$ over the Doppler width $|n|\lesssim\bar\omega R\sin i$. The lens leaves this width unchanged and redistributes power among
the lines [Fig.~\ref{fig:bound}(b)]. Since the comb and Eq.~\eqref{eq:hT} are the same convolution evaluated in frequency and in time,
their comparison tests the whole construction: they agree to $3.2\times10^{-5}$ in modulus and $3\times10^{-5}$\,rad in phase, three
orders of magnitude below the adiabatic error for the same configuration, so that the check resolves the retardation of individual
worldlines.

The observable is the lensing modulation
\begin{equation}\label{eq:Lmod}
\Lc(T)=\frac{h(T)}{h_0(T)},
\end{equation}
the ratio of the lensed signal to the unlensed one~\eqref{eq:h0}. Figure~\ref{fig:bound} shows it for $w=1$, $R=100M$ and $\cos i=0.1$,
together with the adiabatic and eikonal models. Not all of $\Lc$ is measurable: its lowest Fourier harmonics in the orbital phase,
$|n|\le2$, are degenerate with the parameters of an unlensed fit of the orbit (R\o mer delay, Doppler shift and Shapiro delay), and we
remove them. The rms over one orbit of what remains defines the non-degenerate lensing signal, $Q_\phi$ in phase and $Q_{\ln A}$ in
log-amplitude. Finally, the content beyond the Shapiro delay is measured by $\Lc/\Lc_{\rm eik}$, whose rms phase we denote
$Q_\phi^{\rm beyond\,Sh}$; by construction, it is also the error of the eikonal model.

\section{Results}\label{sec:results}

With the exact signal at hand we can now measure how well each approximate description reproduces it, and where along the orbit
each of them breaks down. Besides the adiabatic model~\eqref{eq:adiab} we consider the eikonal model, the paraxial model and two
forms of geometric optics. Geometric optics does not require the adiabatic step. The delay density~\eqref{eq:rho_std} is singular at
the arrival of each image, with a step at the minimum and a logarithm at the saddle, and keeping only these singularities in
Eq.~\eqref{eq:hT} gives the time-domain form of geometric optics,
\begin{equation}\label{eq:hGO}
h_{\rm GO}(T)=\sqrt{\mu_+(t_+)}\,q(t_+)\,\frac{\dd t_+}{\dd T}-\sqrt{\mu_-(t_-)}\;\mathcal H\Big[q(t_-)\frac{\dd t_-}{\dd T}\Big],
\end{equation}
where the retarded times $t_\pm$ solve $T=t_\pm+\Delta_R(t_\pm)+4M\,\Tf_\pm\big(y(t_\pm)\big)$, with $\Tf_\pm$ the Fermat delays of the
two images, and $\mathcal H$ is the Hilbert transform, which for a monochromatic signal is the Morse phase $-\ii$. In this form each
classical worldline has its own retarded time, its own magnification evaluated at that time and its own Doppler factor. Far from
conjunction the secondary worldline is the echo, with delay $4M\Tf_-\simeq\eta+2M\ln(\eta/2M)+4M$. The other models follow from
further simplifications. Geometric optics at one retarded time evaluates both classical worldlines at the direct retarded time and
at the local frequency; the paraxial model is the adiabatic model with $\eta_{\rm par}$ in place of $\eta$; and the eikonal model keeps
a single straight worldline at the direct retarded time, $\Lc_{\rm eik}=y^{-\ii w_{\rm loc}}$, i.e.\ the Shapiro delay at the local
frequency. We measure the error of each model by the rms of its non-degenerate phase error over the orbit, defined as $Q_\phi$ in
Sec.~\ref{sec:comb}.

\subsection{The exact modulation}\label{sec:bound_exact}

The exact modulation for $w=1$, $R=100M$ and $\cos i=0.1$ is shown in Fig.~\ref{fig:bound}. It has two distinct components: the
flare at conjunction, where $\arg\Lc$ reaches $1.6$\,rad and $\ln|\Lc|$ reaches $0.5$, and, on the rest of the orbit, the
oscillations produced by the echo as it interferes with the direct wave. Table~\ref{tab:results} lists the non-degenerate
modulation and the errors of the models for a subset of the configurations. The phase modulation
grows roughly linearly with $w$, from $Q_\phi=0.03$--$0.08$ at $w=0.3$ to 1--3\,rad at $w=10$. Most of it, however, is the Shapiro
delay, which an orbital timing model would absorb, and the content beyond the Shapiro delay is 8--60\% of $Q_\phi$ for $w\ge1$. The
amplitude modulation is $0.02$--$0.11$ and depends only weakly on $w$. The relevant benchmark for the approximate models is
therefore the part of the signal beyond the Shapiro delay, and the models fail in different ways near conjunction and on the far
side.

\begin{table}[h!]\centering\small\setlength{\tabcolsep}{4.5pt}
\begin{tabular}{rrr|ccc|cccc}
\toprule
$w$ & $R/M$ & $\cos i$ & $Q_\phi$ & $Q_{\ln A}$ & $Q_\phi^{\rm beyond\,Sh}$ & paraxial & GO, one $t_{\rm ret}$ & GO, own $t_{\rm ret}$ & adiabatic\\
\midrule
0.3 & 100 & 0.1 & 0.030 & 0.044 & 0.071 & $1.2{\times}10^{-2}$ & $3.9{\times}10^{-2}$ & $3.8{\times}10^{-2}$ & $8.4{\times}10^{-3}$\\
1 & 100 & 0.1 & 0.186 & 0.106 & 0.111 & $2.4{\times}10^{-2}$ & $2.4{\times}10^{-2}$ & $1.1{\times}10^{-2}$ & $2.2{\times}10^{-2}$\\
3 & 100 & 0.1 & 0.578 & 0.096 & 0.202 & $5.8{\times}10^{-2}$ & $3.6{\times}10^{-2}$ & $8.9{\times}10^{-3}$ & $3.5{\times}10^{-2}$\\
10 & 100 & 0.1 & 1.750 & 0.105 & 0.804 & $1.7{\times}10^{-1}$ & $6.5{\times}10^{-2}$ & $5.4{\times}10^{-3}$ & $6.5{\times}10^{-2}$\\
\midrule
0.3 & 1000 & 0.02 & 0.079 & 0.036 & 0.054 & $6.4{\times}10^{-3}$ & $3.2{\times}10^{-2}$ & $3.2{\times}10^{-2}$ & $2.0{\times}10^{-3}$\\
1 & 1000 & 0.02 & 0.310 & 0.071 & 0.104 & $2.0{\times}10^{-2}$ & $8.4{\times}10^{-3}$ & $7.7{\times}10^{-3}$ & $3.5{\times}10^{-3}$\\
3 & 1000 & 0.02 & 0.946 & 0.080 & 0.222 & $5.9{\times}10^{-2}$ & $7.8{\times}10^{-3}$ & $4.9{\times}10^{-3}$ & $6.0{\times}10^{-3}$\\
10 & 1000 & 0.02 & 3.082 & 0.078 & 0.830 & $2.0{\times}10^{-1}$ & $1.2{\times}10^{-2}$ & $2.2{\times}10^{-3}$ & $1.1{\times}10^{-2}$\\
\bottomrule
\end{tabular}
\caption{Bound lens over one orbit . Columns 4--6: rms of the non-degenerate phase (rad) and
log-amplitude of the exact modulation, and the phase content beyond the Shapiro delay (equal to the error of the eikonal model).
Columns 7--10: rms non-degenerate phase error (rad) of the approximate models.}\label{tab:results}
\end{table}

\subsection{Near conjunction: diffraction}\label{sec:bound_conj}

\begin{figure}[h!]\centering
\includegraphics[width=\textwidth]{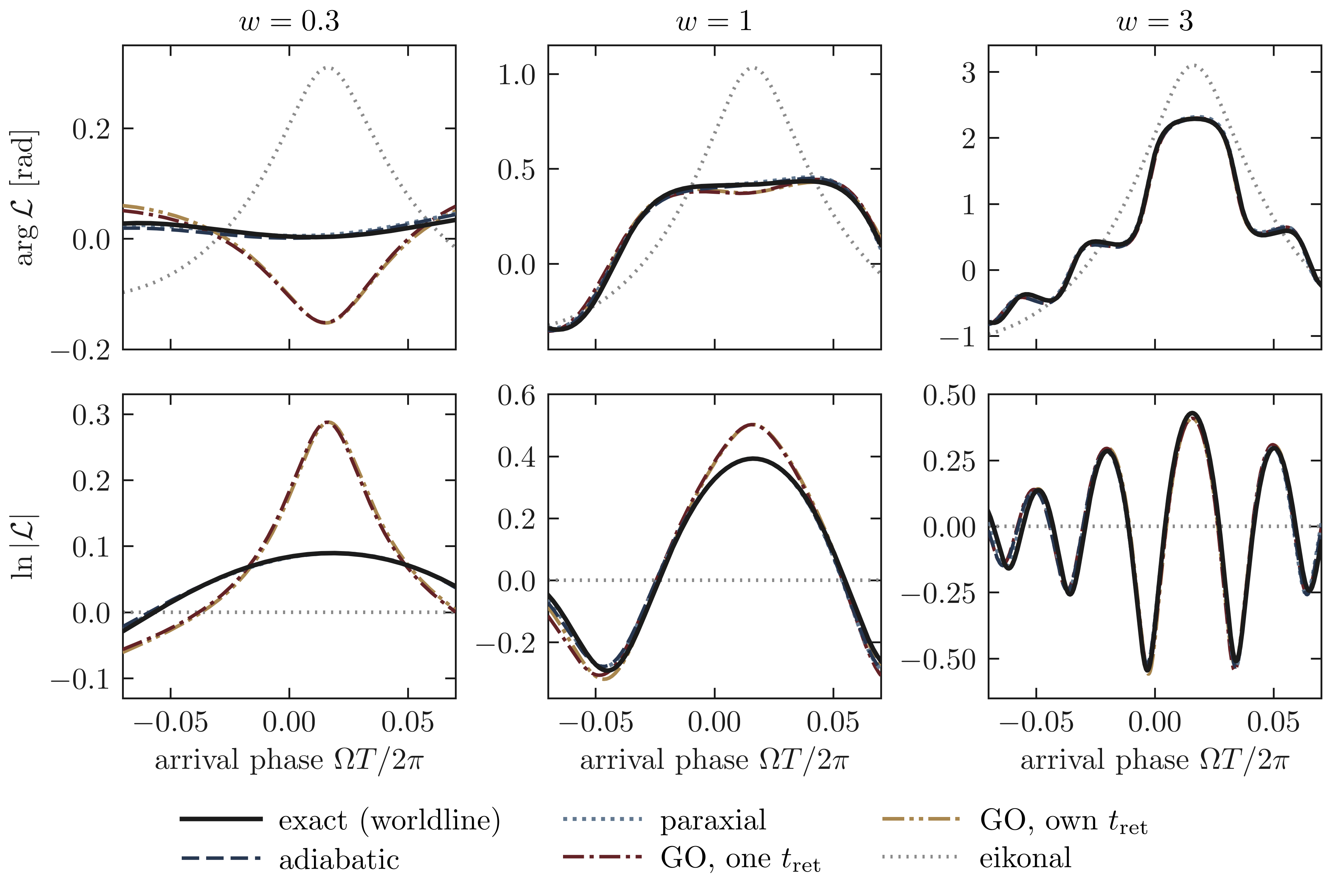}
\caption{Near conjunction: the exact modulation (black) and the five models, for $R=100M$, $\cos i=0.1$ and $w=0.3$, 1 and 3. Top: $\arg\Lc$; bottom: $\ln|\Lc|$; harmonics $|n|\le2$ removed. At $w=0.3$ the two GO
models predict a phase dip where the exact phase is flat and an amplitude flare three times too large. The eikonal is the
Shapiro delay alone.}\label{fig:models}
\end{figure}

Near conjunction the lensing signal beyond the Shapiro delay is 0.044--0.21\,rad at $w=0.3$ and 0.10--0.40\,rad at $w=1$, and this is
where geometric optics fails. At $w=0.3$ both GO models are in error by 0.026--0.12\,rad in phase and by up to 0.15 in log-amplitude,
that is, by 46--100\% of the signal. Geometric optics predicts a phase dip where the exact phase is flat, and an amplitude flare three
times too large (Fig.~\ref{fig:models}). At $w=1$ the error is smaller but still substantial, 0.015--0.029\,rad in phase (7--19\% of
the signal) and 0.016--0.079 in log-amplitude (17--28\%). Giving each image its own retarded time changes little here, because the
error has nothing to do with the motion of the source. It comes from the smooth part of the delay density, which geometric optics
discards, and it is present for a static source as well: $|F_{\rm GO}/F-1|=0.15$, 0.04 and 0.02 at $w=1$ and $y\simeq0.5$, 1 and 3. Within
$\beta_{\rm GO}$ of conjunction, Eq.~\eqref{eq:betaGO}, the wave does not separate into two classical worldlines, and no choice of
retarded times can repair this.

The adiabatic model keeps the exact static amplification factor, and with it all the diffraction, and is the most accurate
approximation near conjunction for $w\lesssim1$. In the conjunction window its error at $w=1$ is $1.4$--$1.7\times10^{-2}$\,rad for
$R=100M$ and $1.5$--$1.9\times10^{-3}$\,rad for $R=10^3M$, i.e.\ 7--13\% and 0.5--1.4\% of the lensing signal beyond the Shapiro delay.
What remains is the transit effect of Sec.~\ref{sec:bound_models}: the worldlines of the diffraction spot have different delays and
see the source at slightly different offsets, an error of order $wM/R$.

\subsection{The far side: geometry and the echo}\label{sec:bound_farside}

On the far side diffraction is no longer the issue, and the models fail through the geometry and the timing. The paraxial model
evaluates the Shapiro phase with the wrong geometry, $\nu\ln(\eta/\eta_{\rm par})$, and its error therefore grows linearly with $w$,
from 0.007--0.012\,rad at $w=0.3$ to 0.13--0.25\,rad at $w=10$. The adiabatic model has the right geometry but the wrong timing of the
echo. For $w\ge1$ its far-side error is $(1.5$--$1.7)\times10^{-2}$\,rad at $R=100M$ and $(1.2$--$1.6)\times10^{-3}$\,rad at $R=10^3M$,
nearly independent of $w$ and equal to the echo amplitude: the adiabatic echo carries the Doppler factor of the direct wave, so
that its phase drifts away from that of the true echo along the orbit (Sec.~\ref{sec:bound_models}).

\begin{figure}[h!]\centering
\includegraphics[width=\textwidth]{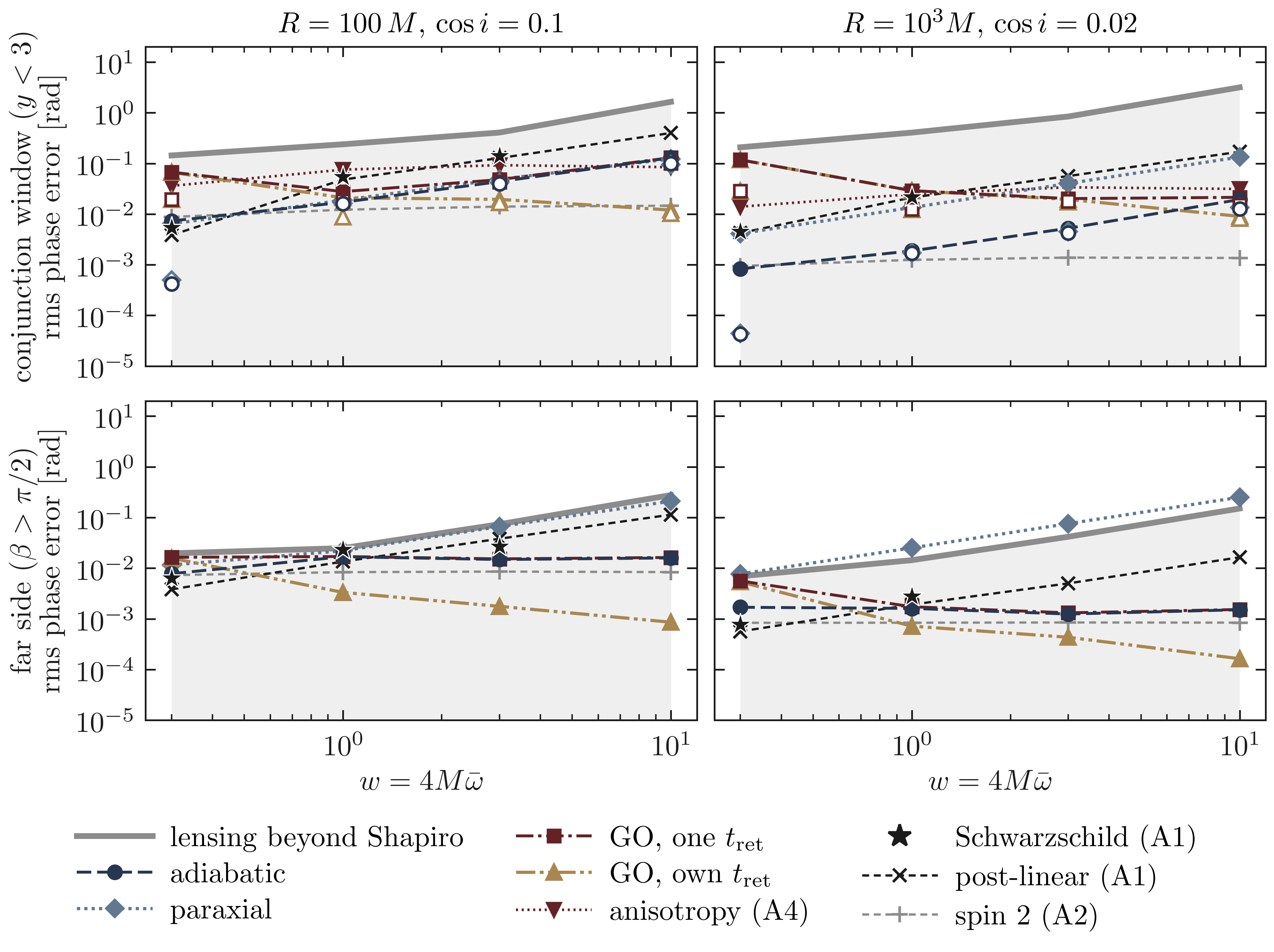}
\caption{Error budget against $w$: rms non-degenerate phase errors of four models
(coloured; filled: whole-orbit fit of the degenerate harmonics; open: conjunction window with a window-local fit) and the
corrections to the scalar linearized model with an isotropic source (black stars: exact Schwarzschild; grey dashed: first-order
post-linear form; blue dotted: anisotropy of the emission with $|\bm g|=1$). The shaded region lies below the lensing content
beyond the Shapiro delay, which equals the eikonal error. Top: conjunction window ($y<3$); bottom: far side ($\beta>\pi/2$). Left:
$R=100M$, $\cos i=0.1$; right: $R=10^3M$, $\cos i=0.02$.}\label{fig:budget}
\end{figure}

Geometric optics at the images' own retarded times, on the contrary, delivers the echo at the right time and with the right
Doppler factor, and on the far side the echo is a well separated classical worldline. For $w\ge1$ this is the most accurate model
there, with errors of $3.8\times10^{-5}$--$3.3\times10^{-3}$\,rad, and over the whole orbit at $w\ge3$ it reproduces the exact result
to $3.4\times10^{-4}$--$8.9\times10^{-3}$\,rad. The two parts of the orbit thus probe different physics (Fig.~\ref{fig:budget}). For
$w\gtrsim3$ what the exact result adds to the adiabatic product is the retardation of each worldline, which geometric optics with
the images' own retarded times captures. For $w\lesssim1$ there are no separate worldlines to retard, and only the exact formula is
accurate at the level of the transit effect.

\subsection{Beyond the scalar linearized model}\label{sec:bound_beyond}

All the results above assume the model of Sec.~\ref{sec:bound_model}: a scalar wave in the linearized field of a static lens,
emitted isotropically. Relaxing these assumptions does not change the structure of the calculation. The corrections enter only
through the static inputs of the time-domain formula, the amplification factor (or the delay density $\rh$) and the source
function $q$, while the kinematics of Eq.~\eqref{eq:hT}, with its retarded times and Doppler factors, is unchanged. It is therefore
enough to estimate how much each correction changes the static inputs, and to compare it with the lensing signal and with the
errors of the models.

The largest correction comes from the field of the lens itself. On the exact Schwarzschild metric the scalar wave of a point
source at finite distance separates into partial waves that obey the Regge--Wheeler equation with $s=0$ \cite{ReggeWheeler1957},
and the sum converges absolutely, because for $\ell>\omega R$ the source lies in the classically forbidden region of the radial
problem. Near conjunction the correction is dominated by the second-order deflection, $\Theta=4M/b+15\pi M^2/4b^2$, which adds to
the Fermat potential the term $\epsilon/|\bx|$ with $\epsilon=\frac{15\pi}{32}(M/R)^{1/2}$ (0.147 at $R=100M$) and gives a closed-form
first-order correction. This reproduces the exact correction with a residual of 6--14\% at $R=10^3M$ and 20--50\% at $R=100M$. On the
far side the two differ by factors 0.5--2.3. There the echo comes from $b\le2M$ in the linearized model, whereas in Schwarzschild large
deflections arise only for $b\to3\sqrt3M^+$, and the horizon absorbs the lowest partial waves; the far-side echo is a strong-field
effect, as anticipated in Sec.~\ref{sec:bound_far}. A rotating (Kerr) lens adds a term $\simeq0.34\,a/M$ times the post-linear one,
largely degenerate with the orbit; strong-field lensing by a Kerr black hole has been computed in Ref.~\cite{Saketh:2025cwf}.

The spin of the wave matters less. In the long-wavelength regime the helicity-preserving and helicity-reversing scattering
amplitudes of GWs on Schwarzschild are $f_s\cos^4(\theta/2)$ and $f_s\sin^4(\theta/2)$, with $f_s$ the scalar amplitude
\cite{Westervelt1971,Peters1976,DeLogiKovacs1977,Dolan2008a}. For the bound lens the scattering angle is $\beta$, and the
helicity-preserving signal differs from the scalar one by $|\delta F_{++}|=(1+\cos^2\frac\beta2)M/R$, a correction $\simeq M/R$ that
is nearly independent of $w$.

The emission pattern of the source enters in a different way. Each worldline leaves the source in its own direction, so the two
images see the binary from slightly different directions. With $\bm g$ the complex sky gradient of the logarithm of the emission
amplitude, the first-order correction has a closed form. For the dominant quadrupole helicity $|\bm g|=2\sqrt2\tan(\iota/2)$, where
$\iota$ is the inclination of the binary, from 0 for a face-on binary to 2.8 for an edge-on one, and for nearly edge-on binaries this
is the largest correction near conjunction. Finally, the reflex motion of the lens is of relative order $(m/M)v$.

For $w\gtrsim1$ and $R=10^2$--$10^3M$ the corrections are therefore ordered as
\begin{equation}
\underbrace{w\Big(\frac MR\Big)^{1/2},\ |\bm g|\,w\Big(\frac MR\Big)^{1/2}}_{\rm Schwarzschild,\ anisotropy}
>\underbrace{w\,\frac MR}_{\rm transit}\sim\underbrace{\frac MR}_{\rm spin\,2,\ echo},
\end{equation}
and Table~\ref{tab:hierarchy} collects the budget at $w=1$. None of them removes the inadequacy of geometric optics and of the
eikonal in the diffraction regime. Near conjunction the lensing content beyond the Shapiro delay, 0.04--0.4\,rad for $w\le1$, is at
least $\simeq1.5$ times larger than every computed correction for $|\bm g|\lesssim1$, and at $w\simeq0.3$ the GO error near conjunction
is 4.5--26 times the Schwarzschild correction.

\begin{table}[h!]\centering\small\setlength{\tabcolsep}{4pt}
\begin{tabular}{l>{\raggedright\arraybackslash}p{3.1cm}cccc}
\toprule
Effect & Scaling & \multicolumn{2}{c}{$R=100M$, $\cos i=0.1$} & \multicolumn{2}{c}{$R=10^3M$, $\cos i=0.02$}\\
 & & conjunction & far side & conjunction & far side\\
\midrule
paraxial geometry & $\propto w$ (far side) & $1.9\times10^{-2}$ & $2.2\times10^{-2}$ & $1.3\times10^{-2}$ & $2.5\times10^{-2}$\\
GO, one $t_{\rm ret}$ & $wy\lesssim1$ & $2.8\times10^{-2}$ & $1.7\times10^{-2}$ & $2.9\times10^{-2}$ & $1.7\times10^{-3}$\\
GO, own $t_{\rm ret}$ & $wy\lesssim1$ & $2.1\times10^{-2}$ & $3.3\times10^{-3}$ & $2.9\times10^{-2}$ & $7.2\times10^{-4}$\\
adiabatic & $(1$--$2)wM/R$; echo & $1.7\times10^{-2}$ & $1.7\times10^{-2}$ & $1.9\times10^{-3}$ & $1.6\times10^{-3}$\\
\midrule
Schwarzschild & $\sim w(M/R)^{1/2}$ & $5.4\times10^{-2}$ & $2.3\times10^{-2}$ & $2.2\times10^{-2}$ & $2.8\times10^{-3}$\\
spin 2 & $\simeq M/R$ & $1.2\times10^{-2}$ & $8.3\times10^{-3}$ & $1.2\times10^{-3}$ & $8.4\times10^{-4}$\\
anisotropy, per $|\bm g|$ & $\sim|\bm g|w(M/R)^{1/2}$ & $7.5\times10^{-2}$ & --- & $2.5\times10^{-2}$ & ---\\
\bottomrule
\end{tabular}
\caption{Error budget at $w=1$ (rms non-degenerate phase, rad; whole-orbit fit), . The lensing
content beyond the Shapiro delay, equal to the eikonal error, is 0.24 and 0.025\,rad in the two windows for $R=100M$, and 0.40 and
0.014\,rad for $R=10^3M$.}\label{tab:hierarchy}
\end{table}

\subsection{Chirping sources, distinguishability and the astrophysical window}\label{sec:bound_obs}

\begin{figure}[h]\centering
\includegraphics[width=\textwidth]{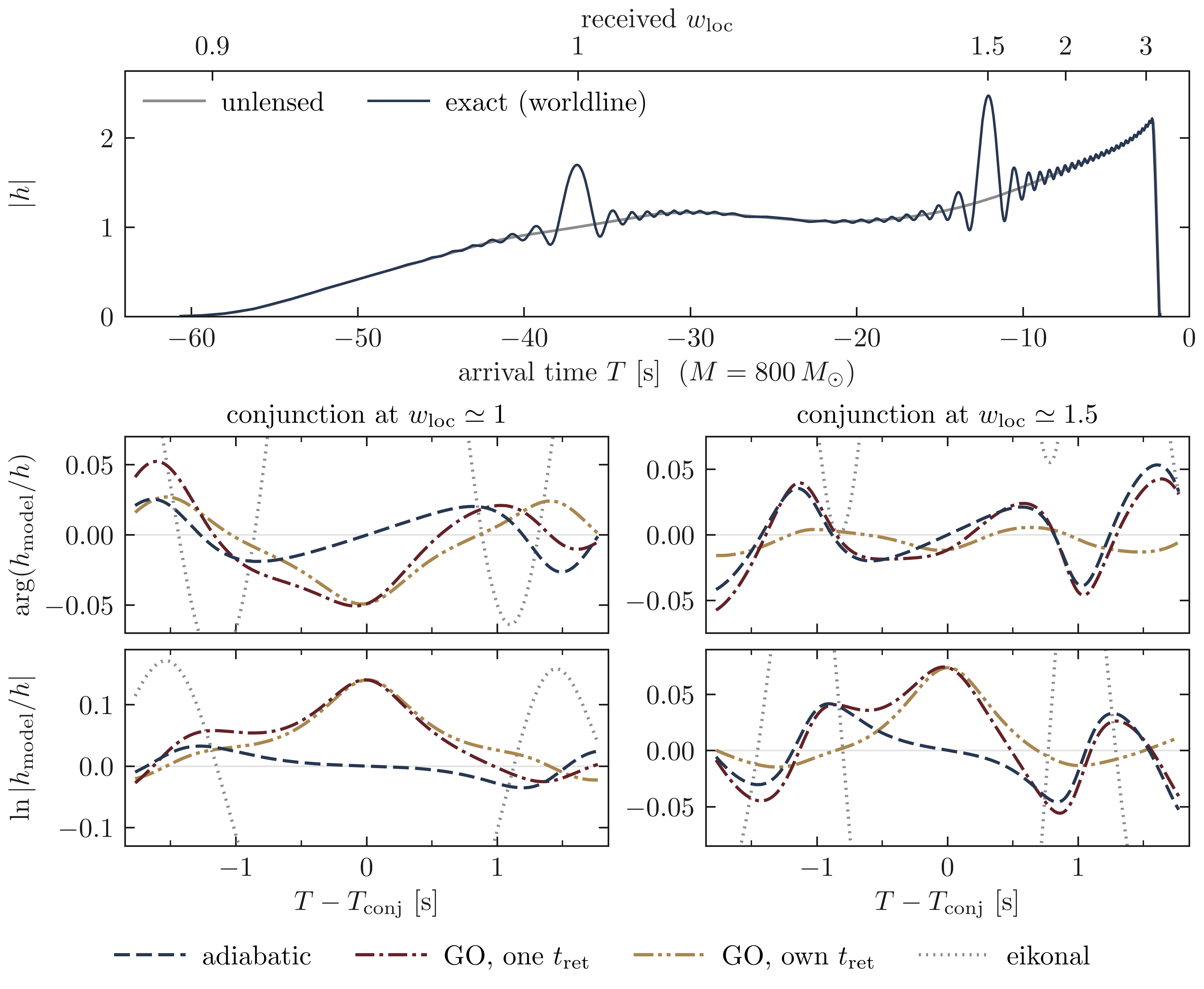}
\caption{Chirping source: a $10+10\,M_\odot$ inspiral orbiting an $800M_\odot$ black hole at $R=100M$, with
$\cos i=0.1$. The source waveform covers the band $0.8\le w\le3.5$ ($f\simeq8$--$35$\,Hz) and is switched on and
off smoothly at its edges, so $|h|\to0$ at both ends. Top: $|h|$ of the exact (worldline) and unlensed signals;
the top axis gives the received $w_{\rm loc}$, and the two peaks are the conjunctions at $w_{\rm loc}\simeq1$ and
$1.5$. Bottom: errors of the approximate models near these conjunctions, i.e.\ phase and log-amplitude of
$h_{\rm model}/h$ without any fit, so that zero is the exact result ($1\,{\rm s}\simeq250M$). }\label{fig:chirp}
\end{figure}

The time-domain formula~\eqref{eq:hT} holds for any $q(t)$, and its main application is to sources whose frequency sweeps through
the diffraction regime while they move along the orbit. Consider a $10+10\,M_\odot$ inspiral orbiting an $800M_\odot$ black hole, for
which $w=1$ at $f\simeq10$\,Hz. In the band $0.8\le w\le3.5$, i.e.\ $f\simeq8$--$35$\,Hz and 65\,s of signal, it covers 2.6 orbits at
$R=100M$ (Fig.~\ref{fig:chirp}). At the conjunction with $w_{\rm loc}\simeq1$ and $y_{\min}=0.5$ both GO models overestimate the
amplitude by 15\%. The adiabatic model is exact there to $3\times10^{-4}$, but a few hundred $M$ before and after conjunction the
transit effect gives errors up to $(3$--$5)\times10^{-2}$, and away from conjunction the model misplaces the echo. A single chirping
signal thus displays the failures found above for monochromatic sources one after the other.

\begin{figure}[h]\centering
\includegraphics[width=0.9\textwidth]{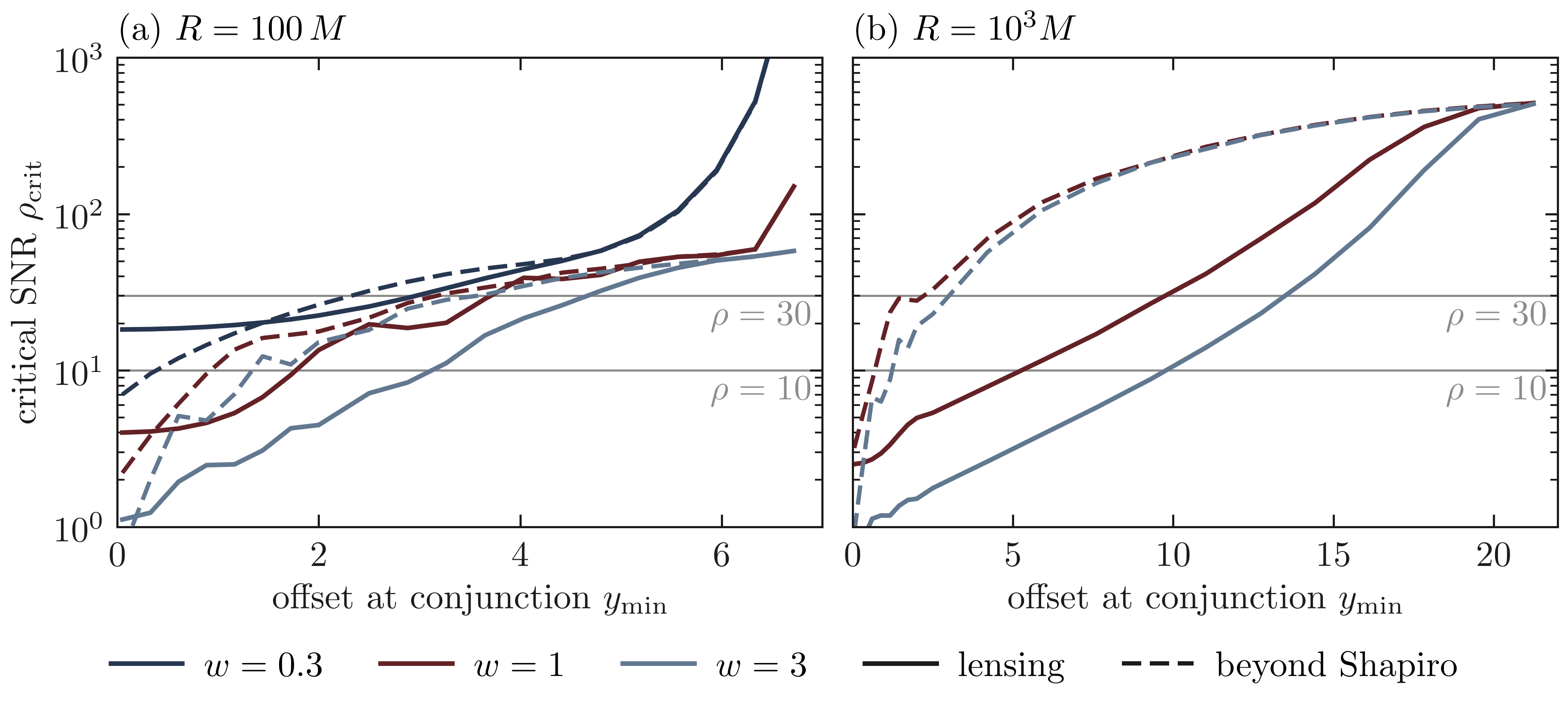}
\caption{Critical signal-to-noise ratio $\rho_{\rm crit}$ above which the exact signal can be told apart from a simpler
one, against the offset at conjunction $y_{\min}=[R(1-\sin i)/2M]^{1/2}\simeq(R/4M)^{1/2}|\cos i|$, which runs from
edge-on orbits ($y_{\min}=0$) to nearly face-on ones (right end of each axis); left $R=100M$, right $R=10^3M$. Solid:
exact versus unlensed signal, i.e.\ whether lensing is detectable at all; dashed: exact versus eikonal signal, i.e.\
whether the lensing beyond the Shapiro delay (magnification, second image and diffraction) is detectable. Two signals
are distinguishable for $\rho>\rho_{\rm crit}=(2\mathcal M_\perp)^{-1/2}$, with $\mathcal M_\perp$ the mismatch after
projecting out the directions degenerate with the orbital parameters, for a quasi-monochromatic source observed over
whole orbits in white noise; lower curves mean easier detection, and values $\lesssim3$ only indicate a large
difference. Grey lines: $\rho=10$ and 30. All curves rise with $y_{\min}$, as the source passes further from the
lens--observer axis, and fall as $w$ grows. At $R=100M$ and the smallest offsets ($y_{\min}\lesssim1.5$ for $w=0.3$)
the dashed curves lie below the solid ones: there the eikonal overshoots, since its Shapiro phase $-w\ln y$ grows
without bound at conjunction while diffraction keeps the exact phase finite. At large offsets the two curves merge,
because the Shapiro modulation becomes a smooth function of the orbital phase and is absorbed by the orbital
parameters. At $R=10^3M$ lensing is detectable at $\rho\le10$ up to $y_{\min}\simeq5$ ($w=1$) and 10 ($w=3$), but
the content beyond the Shapiro delay only for $y_{\min}\lesssim1$, because it is concentrated in the conjunction window,
a small fraction of a long orbit.}\label{fig:detect}
\end{figure}

Whether these differences can be measured depends on the signal-to-noise ratio. Two signals are distinguishable when their
optimally aligned difference has $\langle\delta h|\delta h\rangle>1$ \cite{Lindblom2008}, i.e.\ above the signal-to-noise ratio
$\rho_{\rm crit}=(2\mathcal M_\perp)^{-1/2}$, where $\mathcal M_\perp$ is the mismatch after projecting out the directions degenerate with
the orbit. For $w=1$--3 and $y_{\min}\lesssim1.5$, lensing is distinguishable from an unlensed signal at $\rho\simeq2$--7, and lensing
beyond the Shapiro delay at $\rho\simeq4$--16. At $R=100M$ and $w\ge1$ the adiabatic model, geometric optics with one retarded time and
the Schwarzschild correction become visible at $\rho\simeq10$--35. At $R=10^3M$ the adiabatic model remains adequate up to
$\rho\gtrsim110$, whereas geometric optics fails at $\rho\simeq15$--30 for small offsets and $w\lesssim1.5$ (Table~\ref{tab:detect},
Fig.~\ref{fig:detect}). The lens is therefore detectable at modest signal-to-noise ratios, and its diffraction beyond the Shapiro
delay at only slightly larger ones.

\begin{table}[h!]\centering\small
\begin{tabular}{lccccccc}
\toprule
$R/M$ & 100 & 100 & 100 & 100 & 1000 & 1000 & 1000\\
$\cos i$ & 0.1 & 0.1 & 0.1 & 0.3 & 0.02 & 0.02 & 0.1\\
$w$ & 0.3 & 1 & 3 & 1 & 0.3 & 1 & 1\\
\midrule
lensing (vs unlensed) & 19 & 4 & 1.7 & 7 & 11 & 3 & 4\\
beyond Shapiro (vs eikonal) & 12 & 7 & 5 & 16 & 15 & 7 & 28\\
GO, one $t_{\rm ret}$ & 15 & 23 & 20 & 32 & 14 & 27 & 124\\
GO, own $t_{\rm ret}$ & 15 & 32 & 92 & 84 & 14 & 27 & 154\\
adiabatic & 85 & 34 & 21 & 34 & 359 & 205 & 203\\
Schwarzschild & 93 & 24 & 10 & 26 & 449 & 105 & 159\\
spin 2 & 73 & 70 & 67 & 66 & 664 & 663 & 649\\
\bottomrule
\end{tabular}
\caption{Critical signal-to-noise ratio $\rho_{\rm crit}$ above which the exact signal is distinguishable from the given model
(quasi-monochromatic source over whole orbits, white noise). Values $\lesssim3$ only indicate
a large difference.}\label{tab:detect}
\end{table}

\begin{figure}[h!]\centering
\includegraphics[width=0.62\textwidth]{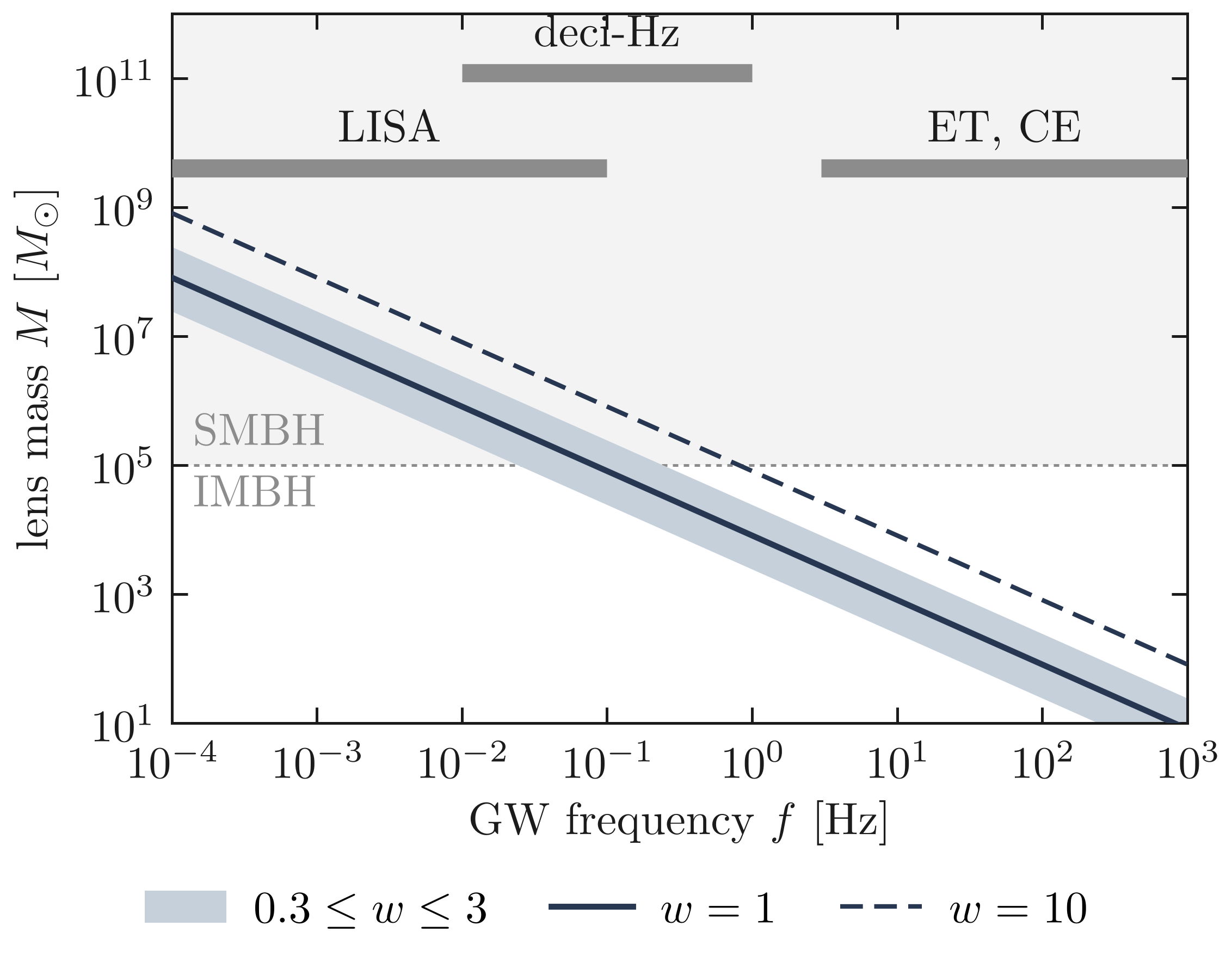}
\caption{The diffraction regime $0.3\le w\le3$ (blue band) in the plane of GW frequency and lens mass, with the bands of LISA,
deci-Hz and third-generation detectors.}\label{fig:window}
\end{figure}

It remains to see where such systems can be found. Since $w\simeq1.24\times10^{-4}(M_z/M_\odot)(f/{\rm Hz})$, with $M_z$ the redshifted
mass of the lens, the diffraction regime $0.3\le w\le3$ corresponds to $M\simeq(0.24$--$2.4)\times10^4M_\odot\,(f/{\rm Hz})^{-1}$
(Fig.~\ref{fig:window}), and Table~\ref{tab:astro} gives the relevant time scales at $w=1$. In the LISA band the lens is a nuclear black
hole like that of the Milky Way, the orbital period at $R=100$--$10^3M$ is 1--30\,d and the maximum echo delay 0.7--7\,h; a
$30+30\,M_\odot$ inner binary at 3\,mHz is quasi-monochromatic over many orbits. For third-generation detectors the lens is an
intermediate-mass black hole of $\sim800M_\odot$, and a chirp lasts less than an orbit. For a source observed over many orbits with
isotropic orientation, the probability that the offset at conjunction is below $y_c$ is $P=2y_c(M/R)^{1/2}$. With the threshold set by
the signal-to-noise ratio, at $w=1$ and $\rho=10$--30, this gives $P\simeq0.2$--0.6 at $R=100M$ and 0.04--0.14 at $R=10^3M$ for lensing
beyond the Shapiro delay. Once a source is in this configuration lensing is therefore likely, and the rates are limited by the poorly
known abundance of compact binaries within $\sim10^3M$ of a massive black hole of the right mass \cite{Sberna2022,Bellovary:2015ifg,Grishin2024}.

\begin{table}[h!]\centering\small
\begin{tabular}{lccc}
\toprule
 & LISA, 3\,mHz & deci-Hz, 0.1\,Hz & 3G, 10\,Hz\\
\midrule
$M$ [$M_\odot$] & $2.7\times10^6$ & $8.1\times10^4$ & 810\\
$P$ ($R=100M$) & 0.97\,d & 42\,min & 25\,s\\
$P$ ($R=10^3M$) & 30\,d & 0.92\,d & 13\,min\\
$2R$ ($R=100M$) & 0.74\,h & 80\,s & 0.8\,s\\
$2R$ ($R=10^3M$) & 7.4\,h & 13\,min & 8\,s\\
\bottomrule
\end{tabular}
\caption{Diffraction-regime examples at $w=1$: lens mass, orbital period $P$ and maximum echo delay $2R$.}\label{tab:astro}
\end{table}

\section{Conclusions}\label{sec:concl}

The main results of this paper are an exact solution of the point-mass lens beyond the diffraction integral, obtained
with the proper-time representation of the propagator, and its application to the bound lens, a configuration in which
the approximations behind the diffraction integral fail together and which therefore serves as a benchmark for the
approximations used in the literature.

For a point mass, the Duru--Kleinert clock, which measures the Shapiro delay accumulated along each worldline, and the
Kustaanheimo--Stiefel map make the proper-time path integral Gaussian and give the propagator in closed form in the
linearized field, at every frequency, for every position of source and observer and for every scattering angle.
Nothing is fixed by hand in this construction: the DK time and the fibre angle are integrated over, and stationarity
with respect to them reproduces the Hamiltonian and the fibre constraints. For an observer at infinity these two
variables span a plane, with the DK time as radius, the fibre angle as azimuth and the Fermat potential as phase, and
the exact far-field amplitude takes the form of the point-lens amplification factor $F(w,y)$ obtained from the
diffraction integral, provided that the offset is defined through the excess path of the route via the lens,
$y^2=\eta/2M$, and not by projecting the source onto a lens plane. This agreement is a property of the Coulomb problem,
whose wave depends on the position of the source only through $\eta$; it is not evidence that the eikonal, paraxial and
thin-lens approximations hold, and for a source close to the lens they do not. The same representation explains why
the images are collinear with the lens, the force being central, while the wave receives contributions from every
azimuth around it, and it shows that geometric optics requires both $w\gg1$ and $wy\gg1$. Evaluated exactly for the
point lens, the eikonal condition reads $\varepsilon_E\simeq\bar\lambda/d+GM/c^2d+\mathcal O(\sin^2\theta/2)$: it is a
lower bound on the frequency, far below every detector band for lenses at cosmological distance, it is unrelated to
geometric optics, and its angular term becomes of order unity for waves scattered through large angles.

In the bound lens all these approximations fail at once in the diffraction regime, and we have computed its signal
exactly for a scalar wave in the linearized field of a static point lens. For a source held fixed, the exact far-field
amplitude keeps the point-lens form at every orbital phase, with the offset given by the exact excess path, whereas the
paraxial geometry, which replaces $\eta$ by its small-angle value, is accurate near conjunction but overestimates $\eta$
by up to a factor $\pi^2/4$ on the far side of the orbit. For a moving source only the lens has to be static: the exact
time-domain signal holds for any orbit and any time dependence of the emission, and each worldline delivers the signal
emitted at its own retarded time, with its own Doppler factor, so that no single amplification factor describes the
received wave. Every approximation used in the literature corresponds to a stationarity condition or to a static
assumption applied to this formula, and each fails in a definite part of the orbit.

The comparison with the exact signal sorts these approximations by the physics they miss. Most of the phase modulation
is the Shapiro delay, which an orbital timing model absorbs, and the lensing beyond it, 8--60\% of the phase modulation
for $w\ge1$, is missed entirely by a single straight worldline carrying the Shapiro phase, the eikonal model in the
sense of high-energy scattering. Near conjunction the wave does not separate into classical worldlines: geometric
optics, which keeps only the singular part of the delay density, misses 46--100\% of the lensing beyond the Shapiro
delay at $w=0.3$ and 7--28\% at $w=1$, already for a static source and whatever retarded times are assigned to the
images, while the static amplification factor evaluated along the orbit retains the diffraction and errs only by the
transit effect, $\approx(1$--$2)\,wM/R$. On the far side the errors are geometric and kinematic: the phase error of the
paraxial geometry grows linearly with $w$, and the adiabatic treatment gives the echo the Doppler factor of the direct
wave, an error equal to the echo amplitude, whereas geometric optics with each image at its own retarded time becomes
the most accurate approximation for $w\ge1$ and reproduces the exact signal over the whole orbit to better than
$10^{-2}$\,rad for $w\ge3$. A chirping source that sweeps through $w\sim1$ displays these failures one after the other
within a single signal. Lensing beyond the Shapiro delay is distinguishable at signal-to-noise ratios $\rho\simeq4$--16
for $w=1$--3 and small offsets at conjunction, so that compact binaries within $10^2$--$10^3M$ of a nuclear black hole
in the LISA band, or of an intermediate-mass black hole for third-generation detectors, are natural targets. Our
estimates of the corrections from the Schwarzschild metric, the spin of the wave and the emission pattern of the source
indicate that they do not remove the failures of geometric optics and of the eikonal model near conjunction; a complete
treatment is left to future work.

Several extensions follow naturally. The partial-wave construction carries over to spin~2 and to Kerr lenses, and a
worldline derivation with spin would use the $\mathcal N=4$ spinning particle or the supersymmetric worldline of
spinning bodies~\cite{Bastianelli:2019,Mogull2021}; whether a gauge of Duru--Kleinert type makes it solvable is open.
For extended lenses no Gaussian gauge is known, but the proper-time representation still provides an exact starting
point from which the approximations can be taken one at a time and their validity judged. The point mass and the bound
lens show how much is lost when they are taken together.
\begin{acknowledgments}
This work was supported by a grant from the Simons Foundation (00017375, RJ). This work was in part supported by the European Research Council (ERC Synergy Grant RedH0T, Grant Agreement No.\@ 101225120). Funded by the European Union. Views and opinions expressed are however those of the author(s) only and do not necessarily reflect those of the European Union or the European Research Council Executive Agency. Neither the European Union nor the granting authority can be held responsible for them.

AI usage: Claude Code was used to assist with coding, computations, and
graphic generation. The ideas, inferences, and results all originate in the (human) brains of
the authors. Claude and ChatGPT were used to check citations, spelling, grammar, and
inconsistencies. The text was written by the authors, who are responsible for the results,
findings, and conclusions of this article.
\end{acknowledgments}

\appendix
\section{The proper-time and Duru--Kleinert calculations}\label{App:A}
\label{app:deriv}\label{app:free}\label{app:KS}\label{app:gauss}

 With
$\hat\bp=-(\ii/\omega)\nabla$ the free kernel is
$\langle\bx|\e^{-\ii\omega\tau\hat\bp^2}|\bx'\rangle=(\omega/4\pi\ii\tau)^{3/2}\e^{\ii\omega d^2/4\tau}$, with $d=|\bx-\bx'|$,
so that, writing $p=\epsilon-\ii\omega$ and $q=\epsilon'-\ii\omega d^2/4$ with $\epsilon,\epsilon'\to0^+$, the proper-time
integral is of the Laplace type
\begin{equation}
\int_0^\infty\dd\tau\;\tau^{-3/2}\,\e^{-p\tau-q/\tau}=\sqrt{\frac\pi q}\;\e^{-2\sqrt{pq}},\qquad{\rm Re}\,p,\ {\rm Re}\,q>0 .
\end{equation}
As the regulators vanish, $pq\to-\omega^2d^2/4$ from below the negative real axis, so that $\sqrt{pq}\to-\ii\omega d/2$ and
$\e^{-2\sqrt{pq}}\to\e^{\ii\omega d}$: the regulator selects the outgoing wave. With $\sqrt{\pi/q}=(2\sqrt\pi/d\sqrt\omega)\,\e^{\ii\pi/4}$
and $(\omega/4\pi\ii)^{3/2}=\omega^{3/2}(4\pi)^{-3/2}\e^{-3\ii\pi/4}$ one finds
\begin{equation}
G_0=\frac\ii\omega\,\omega^{3/2}(4\pi)^{-3/2}\e^{-3\ii\pi/4}\,\frac{2\sqrt\pi}{d\sqrt\omega}\,\e^{\ii\pi/4}\,\e^{\ii\omega d}
=\frac{\e^{\ii\omega d}}{4\pi d},
\end{equation}
which we also checked by direct quadrature at finite regulator. The exponent $\omega(\tau+d^2/4\tau)$ is stationary at
$\tau_*=d/2$, where its second derivative is $4\omega/d$; this sets the relative width $(2\pi/\omega d)^{1/2}$ quoted in
Sec.~\ref{sec:propertime}, while the position of the saddle does not depend on the frequency.

The same projector $zz^\dagger=\frac12(r+\bx\cdot\bm\sigma)$ that defines the KS map gives the overlap of two points,
\begin{equation}\label{eq:overlap}
|z^\dagger z'|^2=z'^\dagger\big(zz^\dagger\big)z'=\tfrac12\big(rr'+\bx\cdot\bx'\big)\equiv\varpi^2 .
\end{equation}
For the Laplacian we use $\nabla_u^2=4\sum_a\partial_{z_a}\partial_{\bar z_a}$ together with $\partial_{\bar z_a}x_i=(\sigma_iz)_a$
and $\partial_{z_a}x_i=(z^\dagger\sigma_i)_a$. Acting on a function $f(\bx(\bu))$ this gives
$\sum_a\partial_{z_a}\partial_{\bar z_a}f={\rm tr}(\sigma_i)\,\partial_if+z^\dagger\sigma_j\sigma_iz\,\partial_i\partial_jf$, where the
first term vanishes and the antisymmetric part of $z^\dagger\sigma_j\sigma_iz=\delta_{ij}r+\ii\epsilon_{jik}x_k$ drops out against
$\partial_i\partial_j$, so that on fibre-independent functions
\begin{equation}\label{eq:lapKS}
\nabla_u^2=4r\,\nabla_x^2\,,
\end{equation}
which is the identity behind Eq.~\eqref{eq:hosc}. The volume elements are related by $\dd^4u=J\,\dd^3x\,\dd\theta$, and since
$SU(2)$ rotates $\bu$ and $\bx$ together $J$ can depend only on $r$. Under the dilation $\bu\to\lambda\bu$ one has
$\dd^4u\to\lambda^4\dd^4u$, $\dd^3x\to\lambda^6\dd^3x$ and $r\to\lambda^2r$, hence $J(\lambda^2r)=\lambda^{-2}J(r)$ and $J=C/r$. Comparing
the volume of the ball $|\bu|^2<L$, which is $\pi^2L^2/2$, with $\int_{r<L}\dd^3x\int_0^{2\pi}\dd\theta\,C/r=4\pi^2CL^2$ fixes $C=1/8$:
\begin{equation}\label{eq:measureKS}
\dd^4u=\frac{\dd^3x\,\dd\theta}{8r}.
\end{equation}

Because the KS map is not one to one, the oscillator Green function $g(\bu,\bu')$ of Eq.~\eqref{eq:gS}, which solves
$\omega^2(\hat h-4M)\,g=\delta^4(\bu-\bu')$, has its source at a single point $\bu'$ of the fibre over $\bx'$ and is not a
function on space. The physical source is instead spread uniformly over the fibre. For a fibre-independent test function
$\varphi_0$ one has
$\int\dd^4u\,\varphi_0\int_0^{2\pi}\dd\theta'\,\delta^4(\bu-\e^{\ii\theta'}\bu')=\int_0^{2\pi}\dd\theta'\,\varphi_0(\bx')=2\pi\varphi_0(\bx')$,
while with the measure~\eqref{eq:measureKS} the same integral reads $\frac{2\pi}8\int\dd^3x\,\varphi_0\,T/r$, with $T$ the
$\theta'$-integral of the $\delta$-function. Therefore
\begin{equation}\label{eq:deltalift}
\int_0^{2\pi}\dd\theta'\,\delta^4\big(\bu-\e^{\ii\theta'}\bu'\big)=8r\,\delta^3(\bx-\bx'),
\end{equation}
and we define
\begin{equation}\label{eq:lift}
G_\omega(\bx,\bx')=\frac18\int_0^{2\pi}\dd\theta'\,g\big(\bu,\e^{\ii\theta'}\bu'\big).
\end{equation}
Since $\hat h$ is invariant under a simultaneous rotation of both arguments,
$g(\e^{\ii\alpha}\bu,\e^{\ii\theta'}\bu')=g(\bu,\e^{\ii(\theta'-\alpha)}\bu')$, the $\theta'$-integral removes any dependence on
the fibre angle $\alpha$ of $\bu$, and Eq.~\eqref{eq:lift} is a function of $\bx$ and $\bx'$ only. Applying $\omega^2(\hat h-4M)$ and
using Eq.~\eqref{eq:deltalift} gives $\omega^2(\hat h-4M)G_\omega=r\,\delta^3(\bx-\bx')$, which by Eq.~\eqref{eq:hosc} is
$r\hat H_\omega G_\omega=r\,\delta^3(\bx-\bx')$, i.e.\ $\hat H_\omega G_\omega=\delta^3(\bx-\bx')$: Eq.~\eqref{eq:lift} is the Green
function of the wave equation exactly. The factor $r'$ of Eq.~\eqref{eq:GDK3} has been absorbed by the measure. In terms of
paths, the $\theta'$-integral sums over the starting points on the fibre over $\bx'$; in quantum terms it projects onto the
states of zero fibre charge, which are the functions on space.

For fixed $\tilde t$ and $\theta'$, the path integral in Eq.~\eqref{eq:GDK4} runs from $\bu''\equiv\e^{\ii\theta'}\bu'$ to $\bu$
with Lagrangian $\dot\bu^2+\bu^2$, whose equation of motion $\ddot\bu=\bu$ has the unique solution
\begin{equation}\label{eq:ucl}
\bu_{\rm cl}(\tilde t')=\frac{\bu''\sinh(\tilde t-\tilde t')+\bu\,\sinh\tilde t'}{\sinh\tilde t},
\end{equation}
unique for every $\tilde t>0$ because $\sinh\tilde t\neq0$: the inverted oscillator has no caustics. On this path the action
reduces to a boundary term,
\begin{equation}\label{eq:Acl}
\Acl(\tilde t,\theta')=\int_0^{\tilde t}\dd\tilde t'\,\big(\dot\bu_{\rm cl}^2+\bu_{\rm cl}^2\big)
=\big[\bu_{\rm cl}\cdot\dot\bu_{\rm cl}\big]_0^{\tilde t}
=(r+r')\coth\tilde t-\frac{2\,\bu\cdot\bu''}{\sinh\tilde t},
\end{equation}
with $|\bu|^2=r$ and $|\bu''|^2=r'$. Writing $\bu=\bu_{\rm cl}+\delta\bu$ with $\delta\bu(0)=\delta\bu(\tilde t)=0$, the linear term
$2[\dot\bu_{\rm cl}\cdot\delta\bu]_0^{\tilde t}-2\int\dd\tilde t'\,\delta\bu\cdot(\ddot\bu_{\rm cl}-\bu_{\rm cl})$ vanishes and the
quadratic remainder does not depend on the endpoints, so the path integral is $\e^{\ii\omega\Acl}$ times a function of
$\tilde t$ alone. That function follows from $(\ii/\omega)\partial_{\tilde t}K=\hat hK$. In one dimension, with
$K_1=a(\tilde t)\,\e^{\ii\omega\Lambda}$, $\Lambda=(u^2+u''^2)c-2uu''/{\rm sh}$, $c=\coth\tilde t$ and ${\rm sh}=\sinh\tilde t$, the
spatial derivatives give $\hat hK_1=[-\frac{\ii}{2\omega}c+(uc-u''/{\rm sh})^2-u^2]K_1$ and the time derivative gives
$(\ii/\omega)\partial_{\tilde t}K_1=[\frac\ii\omega a'/a+(u^2+u''^2-2uu''\cosh\tilde t)/{\rm sh}^2]K_1$. The quadratic terms agree
because $c^2-1={\rm sh}^{-2}$, and the remaining ones require $a'/a=-\frac12\coth\tilde t$, i.e.\ $a\propto{\rm sh}^{-1/2}$. For
$\tilde t\to0$, $K_1$ must reduce to the kernel of $\hat p^2/4$, $(\omega/\pi\ii\tilde t)^{1/2}\e^{\ii\omega(u-u'')^2/\tilde t}$, which
fixes $a=(\omega/\pi\ii\,{\rm sh})^{1/2}$. The product over the four directions is Mehler's kernel continued to imaginary frequency,
\begin{equation}\label{eq:mehler}
\int_{\bu(0)=\bu''}^{\bu(\tilde t)=\bu}\mathcal D\bu\;\e^{\ii\omega\int_0^{\tilde t}\dd\tilde t'(\dot\bu^2+\bu^2)}
=\langle\bu|\e^{-\ii\omega\tilde t\hat h}|\bu''\rangle
=-\frac{\omega^2}{\pi^2\sinh^2\tilde t}\;\e^{\ii\omega\Acl(\tilde t,\theta')} .
\end{equation}
The terms of order $1/\omega$ only fix the prefactor, and the phase is $\omega$ times the classical action exactly. The classical
path is used only as an expansion point: $\tilde t$ and $\theta'$ are still integrated over $(0,\infty)$ and $[0,2\pi)$, and each
pair labels one oscillator orbit that in general violates the constraints of Sec.~\ref{sec:constraints}. This is the difference
with the Schwinger gauge, where keeping the classical path is the WKB approximation and fixing the proper time produces the
diffraction integral~\cite{Braga2024}.

Differentiating Eq.~\eqref{eq:Acl} gives the Hamilton--Jacobi relations~\eqref{eq:HJ}. On the one hand
$-\partial_{\tilde t}\Acl=(r+r'-2\,\bu\cdot\bu''\cosh\tilde t)/\sinh^2\tilde t$; on the other, with
$\dot\bu_{\rm cl}(0)=(\bu-\bu''\cosh\tilde t)/\sinh\tilde t$, the conserved energy
$\Eosc=\dot\bu_{\rm cl}^2-\bu_{\rm cl}^2$ evaluated at $\tilde t'=0$ is the same expression. Similarly, with $z''=\e^{\ii\theta'}z'$
one has $\partial_{\theta'}(\bu\cdot\bu'')=-{\rm Im}(z^\dagger z'')$, while the fibre charge at the start of the orbit is
$J=2\,{\rm Im}(z''^\dagger\dot z_{\rm cl}(0))=-2\,{\rm Im}(z^\dagger z'')/\sinh\tilde t$, so that $\partial_{\theta'}\Acl=-J$. The fibre
angle enters only through $\bu\cdot\bu''$, which as a product of real four-vectors is ${\rm Re}(z^\dagger z'')$; with
Eq.~\eqref{eq:overlap},
\begin{equation}\label{eq:varpi}
\bu\cdot\bu''={\rm Re}\big(\e^{\ii\theta'}z^\dagger z'\big)=\varpi\cos(\theta'+\gamma),\qquad \gamma=\arg(z^\dagger z'),
\end{equation}
and the $\theta'$-integral gives a Bessel function,
\begin{equation}\label{eq:fibreJ0}
\int_0^{2\pi}\dd\theta'\,\exp\Big[-\frac{2\ii\omega\varpi\cos(\theta'+\gamma)}{\sinh\tilde t}\Big]
=2\pi\,J_0\Big(\frac{2\omega\varpi}{\sinh\tilde t}\Big).
\end{equation}
Inserting Eqs.~\eqref{eq:mehler} and~\eqref{eq:fibreJ0} in Eq.~\eqref{eq:GDK4}, the constants combine as
$\frac{\ii}{8\omega}\cdot2\pi\cdot\big(-\frac{\omega^2}{\pi^2}\big)=-\frac{\ii\omega}{4\pi}$, which gives Eq.~\eqref{eq:GDK}.

\section{The far field and the lens plane}\label{App:B}
\label{app:hessian}\label{app:far}

For an observer at $\bx=r\bn$, Eq.~\eqref{eq:overlap} gives without approximation
\begin{equation}\label{eq:varpi_far}
\varpi^2=\tfrac12\big(rr'+\bx\cdot\bx'\big)=\tfrac12\,r\big(r'+\bn\cdot\bx'\big)=\frac{r\eta}{2},
\end{equation}
with $\eta$ of Eq.~\eqref{eq:eta}. For large $r$ the phase $\omega(r+r')\coth\tilde t$ in Eq.~\eqref{eq:GDK} varies rapidly
unless $\coth\tilde t-1\simeq2\e^{-2\tilde t}$ is of order $1/\omega r$, which is also the only region where the phase can be
stationary. We therefore set $\tilde t=\chi+\frac12\ln(2\omega r)$, so that $q\equiv\e^{-2\tilde t}=\e^{-2\chi}/2\omega r$, and keep
$\chi$ fixed as $r\to\infty$; the range of $\chi$ becomes the whole real line. Since $\dd\tilde t=\dd l/2r$ along a path on the
energy shell, the logarithm is the DK time accumulated on the long leg to the observer, and $\chi$ the part accumulated near the
lens. With $\sinh\tilde t=\e^{\tilde t}(1-q)/2$ and $q=\mathcal O(1/\omega r)$, the factors of Eq.~\eqref{eq:GDK} become
\begin{equation}\label{eq:farexp}
\begin{aligned}
\frac{1}{\sinh^2\tilde t}&=\frac{4q}{(1-q)^2}=\frac{2\,\e^{-2\chi}}{\omega r}\Big[1+\mathcal O\Big(\frac1{\omega r}\Big)\Big],\qquad
\e^{\ii w\tilde t}=\e^{\ii w\chi}\,\e^{\ii\nu\ln(2\omega r)},\\
\omega(r+r')\coth\tilde t&=\omega(r+r')\frac{1+q}{1-q}=\omega r+\omega r'+\e^{-2\chi}
+\mathcal O\Big(\frac{r'}{r},\frac1{\omega r}\Big),\\
\frac{2\omega\varpi}{\sinh\tilde t}&=\frac{4\omega\varpi\,q^{1/2}}{1-q}=2\sqrt\zeta\,\e^{-\chi}\Big[1+\mathcal O\Big(\frac1{\omega r}\Big)\Big],
\end{aligned}
\end{equation}
with $\nu=w/2$ and $\zeta=\omega\eta$, so that
\begin{equation}
G_\omega(r\bn,\bx')\to-\frac{\ii}{2\pi r}\,\e^{\ii\omega r+\ii\nu\ln2\omega r}\,\e^{\ii\omega r'}
\int_{-\infty}^\infty\dd\chi\;\e^{-2\chi}\,\e^{\ii(w\chi+\e^{-2\chi})}\,J_0\big(2\sqrt\zeta\,\e^{-\chi}\big).
\end{equation}
Writing $J_0$ back as the fibre integral it came from, $J_0(z)=\frac1{2\pi}\int_0^{2\pi}\dd\theta\,\e^{\ii z\cos\theta}$ with
$\theta=\theta'+\gamma+\pi$, and factoring out the unlensed wave through $\e^{\ii\omega r'}=\e^{\ii\zeta}\e^{-\ii\omega\bn\cdot\bx'}$,
gives Eqs.~\eqref{eq:far} and~\eqref{eq:Psi}. The only limit taken is $r\to\infty$; numerically, the relative difference between
Eqs.~\eqref{eq:GDK} and~\eqref{eq:far} is $7\times10^{-4}$ at $\omega r=10^4$ and $2.5\times10^{-4}$ at $\omega r=3\times10^4$, for
$w\simeq2$ and $\omega r'\simeq4.8$.

After the fibre integral, Eq.~\eqref{eq:far} reads
$\e^{-\ii\zeta}F_\omega=-2\ii\int\dd\chi\,\e^{\ii w\chi}\e^{-2\chi}\e^{\ii\e^{-2\chi}}J_0(2\sqrt\zeta\,\e^{-\chi})$. For $M=0$ the
substitution $s=\e^{-2\chi}$ and $\int_0^\infty\dd s\,\e^{-ps}J_0(2\sqrt{\zeta s})=p^{-1}\e^{-\zeta/p}$ with $p=-\ii+0$ give
$\e^{-\ii\zeta}F_\omega=\e^{-\ii\zeta}$, i.e.\ $F_\omega=1$, as it should. For $M\neq0$ the variable $v=\e^{-\chi}$ turns the integral
into $-2\ii\int_0^\infty\dd v\,v^{1-\ii w}\e^{\ii v^2}J_0(2\sqrt\zeta\,v)$, which is of the Mellin--Bessel form~\cite{DLMF}
\begin{equation}
\int_0^\infty v^{p-1}\e^{-\alpha v^2}J_0(cv)\,\dd v=\frac{\Gamma(p/2)}{2\alpha^{p/2}}\,\hF\Big(\frac p2;1;-\frac{c^2}{4\alpha}\Big)
\end{equation}
with $p=2-\ii w$, $\alpha=-\ii+0$ and $c=2\sqrt\zeta$, so that $-c^2/4\alpha=-\ii\zeta$. Since $-\ii\,(-\ii)^{-1}=1$ and
$(-\ii)^{\ii\nu}=\e^{\pi\nu/2}$, this gives $\e^{-\ii\zeta}F_\omega=\e^{\pi\nu/2}\,\Gamma(1-\ii\nu)\,\hF(1-\ii\nu;1;-\ii\zeta)$, and
Kummer's transformation $\hF(a;b;x)=\e^x\hF(b-a;b;-x)$ removes the factor $\e^{-\ii\zeta}$:
\begin{equation}\label{eq:FomegaB}
F_\omega(\bx')=N_w\,\hF\big(\ii\nu,1;\ii\omega\eta\big),\qquad N_w=\e^{\pi\nu/2}\,\Gamma(1-\ii\nu),
\end{equation}
in agreement with Eq.~\eqref{eq:Fomega}. Evaluated with 40-digit arithmetic, the integral agrees with this closed form to
$3\times10^{-37}$.

With the variables of Eq.~\eqref{eq:xmap}, $\e^{-\chi}=\sqrt{w/2}\,|\bx|$ and $\sqrt\zeta=\sqrt{w/2}\,y$, and since
$\bx\cdot\by=-|\bx|y\cos\theta$, the three terms of the phase~\eqref{eq:Psi} become $\e^{-2\chi}=\frac w2|\bx|^2$,
$2\sqrt\zeta\,\e^{-\chi}\cos\theta=-w\,\bx\cdot\by$ and $w\chi=-w\ln|\bx|-\nu\ln(w/2)$. Completing the square with
$\frac w2y^2=\zeta$,
\begin{equation}\label{eq:Psi_lens}
\Psi=w\Big[\tfrac12|\bx|^2-\bx\cdot\by-\ln|\bx|\Big]-\nu\ln\frac w2=w\,\Tf(\bx,\by)-\zeta-\nu\ln\frac w2 .
\end{equation}
For the measure, $\dd^2x=|\bx|\,\dd|\bx|\,\dd\theta$ and $|\bx|\,\dd|\bx|=-\frac2w\e^{-2\chi}\dd\chi$, where the sign is absorbed by the
orientation, since $\chi$ running from $-\infty$ to $\infty$ covers $|\bx|$ from $\infty$ to $0$; hence
$\e^{-2\chi}\dd\chi\,\dd\theta=\frac w2\,\dd^2x$, and $(\chi,\theta)$ covers the plane once. Inserting these in Eq.~\eqref{eq:far},
\begin{equation}\label{eq:lensplane_app}
\e^{\ii\nu\ln(w/2)}F_\omega=\frac{w}{2\pi\ii}\int\dd^2x\;\e^{\ii w\Tf(\bx,\by)},
\end{equation}
which has the form of the point-lens integral~\eqref{eq:kirchhoff} for $F(w,y)\equiv\e^{\ii\nu\ln(w/2)}F_\omega$. The two
normalizations are consistent: with $(-\ii)^{\ii\nu}=\e^{\pi\nu/2}$, the ratio of Eq.~\eqref{eq:Fwy} to Eq.~\eqref{eq:FomegaB} is
$(-\ii w/2)^{\ii\nu}\e^{-\pi\nu/2}=\e^{\ii\nu\ln(w/2)}$. The coincidence relies on the exact solvability of the Coulomb problem and
on the definition $y^2=\eta/2M$ of the offset; near the axis, $\eta=r'(1-\cos\beta)\simeq r'\beta^2/2$ and $y\simeq r'\beta/\xi_E$ with
$\xi_E=\sqrt{4Mr'}$, the usual offset for an observer at infinity.

Near an image the integral~\eqref{eq:far} is Gaussian. Its amplitude is the prefactor $1/\pi$, times the measure
$\e^{-2\chi}=\frac w2x^2$, times $2\pi\,|\det\partial^2\Psi|^{-1/2}=2\pi[w^2(1+x^2)\,|xy|]^{-1/2}$ from the curvatures.
At the images the radial lens equation gives $|xy|=|x^2-1|$, so that $(1+x^2)|xy|=|x^4-1|$ and the amplitude is
\begin{equation}
\frac{x^2}{\sqrt{|x^4-1|}}=\big|1-x^{-4}\big|^{-1/2}=\sqrt{\mu_\pm}.
\end{equation}
Each positive curvature contributes a phase $\e^{\ii\pi/4}$. At image 1 both are positive and their $\e^{\ii\pi/2}$ cancels the
prefactor $-\ii$, while at image 2 the curvature along the fibre angle is negative and the factor $-\ii$ remains: the Morse phase
of the saddle comes from the fibre angle.
\bibliographystyle{plain}

\end{document}